\documentstyle[preprint,aps]{revtex}
\advance\hsize by 0.0truecm  \hoffset=0.7cm
\begin{document}
\bibliographystyle{prsty}
\newcommand{\C}{\cite}
\newcommand{\beq}{\begin{equation}}
\newcommand{\eeq}{\end{equation}}
\newcommand{\bea}{\begin{eqnarray}}
\newcommand{\eea}{\end{eqnarray}}
\newcommand{\bit}{\begin{itemize}}
\newcommand{\eit}{\end{itemize}}
\newcommand{\ms}{m_{\rm s}}
\newcommand{\MeV}{{\rm MeV}}
\newcommand{\fm}{{\rm fm}}
\newcommand{\alf}{{\bar a}}
\newcommand{\bet}{{\bar b}}
\newcommand{\lb}{\hfil\break }
\newcommand{\qeq}[1]{eq.\ (\ref{#1})}
\newcommand{\dsl}{ \rlap{/}{\partial} }
\newcommand{\pst}{ \rlap{/}{p}  }
\newcommand{\half}{\frac{1}{2}}
\newcommand{\quref}[1]{\cite{bolo:#1}}
\newcommand{\qref}[1]{Ref.\ \cite{bolo:#1}}
\newcommand{\queq}[1]{(\ref{#1})}
\newcommand{\qtab}[1]{Tab. \ref{#1}}
\newcommand{\wu}{\sqrt{3}}
\newcommand{\nn}{\nonumber \\ }
\newcommand{\Y}{\ {\cal Y}}
\newcommand{\Sp}{{\rm Sp\ } }
\newcommand{\Spto}{{\rm Sp_{(to)}\ } }
\newcommand{\Tr}{{\rm Tr\ } }
\newcommand{\tr}{{\rm tr\ } }
\newcommand{\sign}{{\rm sign} }
\newcommand{\linie}{\ \vrule height 14pt depth 7pt \ }
\newcommand{\intT}{\int_{-T/2}^{T/2} }
\newcommand{\Nc}{N_{\rm c}}
\newcommand{\Ne}{$N_{\rm c}$}
\newcommand{\gs}{$g_{\rm A}^{(0)}$}
\newcommand{\gt}{$g_{\rm A}^{(3)}$}
\newcommand{\go}{$g_{\rm A}^{(8)}$}
\newcommand{\Gs}{g_{\rm A}^{(0)}}
\newcommand{\Gt}{g_{\rm A}^{(3)}}
\newcommand{\Go}{g_{\rm A}^{(8)}}
\preprint{RUB-TPII-41/93, hep-ph/9403314}

\title{
Axial properties of the nucleon with $1/\Nc$ corrections      \\
   in the solitonic SU(3)-NJL-model  }

\author{
Andree Blotz$^{(1,2)}$
\footnote{email:andreeb@elektron.tp2.ruhr-uni-bochum.de},
Micha{\l} Prasza{\l}owicz$^{(3)}$
\footnote{email:michal@thrisc.if.uj.edu.pl}
and Klaus Goeke$^{(2)}$
\footnote{email:goeke@hadron.tp2.ruhr-uni-bochum.de}  }
\bigskip
\bigskip
\address{(1) Institute for Nuclear Theory (INT), HN-12 University of
Washington,\\
Seattle, WA 98195, USA \\}
\bigskip
\address{(2)
Institute for  Theoretical  Physics  II, \\  P.O. Box 102148,
Ruhr-University Bochum, \\
 D-W-44780 Bochum, Germany  \\
       }
\bigskip
\bigskip
\address{(3)  Institute of Physics,  \\
Jagellonian University, Reymonta 4,30-059, \\  Krakow, Poland}
\bigskip
\date{\today}
\maketitle
\begin{abstract}
Within the semibosonized SU(3)-NJL model  the  mass splittings
of baryons and the
axial vector coupling constants of the nucleon
are evaluated.
The mass splittings of the hyperons are reproduced with great accuracy
if second order corrections in $m_s$ are taken into account.

New corrections to  the axial vector currents coming from the subleading
terms in the $1/\Nc$ expansion are shown to be non-vanishing and
substantially improving the phenomenological predictions for axial
quantities. These corrections are shown to come from two distinctive
sources:
1) anomalous part of the Euclidean effective action related to the
Wess-Zumino term of the SU(3) Skyrme model and 2) real, non-anomalous
part which in this order of $1/\Nc$
has no counterpart within any local
effective meson theory.  The appearance of the type 2) terms
is due to some explicit {\it time-ordering}
of the collective operators entering the formulae for the axial constants.
The question of regularization of these quantities is discussed.
The analytic symmetry breaking terms in the strange quark mass
play a minor
role for $g_{\rm A}^{(3)}$ and $g_{\rm A}^{(0)}$.  They are however
important
for $g_{\rm A}^{(8)}$.
Finally the numerical values for the $g_{\rm A}$'s are
$g_A^{(0)}=0.37$, $g_A^{(3)}=1.38$ and $g_A^{(8)}=0.31$ reproducing
reasonably well the recent data from lepton scattering.
\end{abstract}
\vfill\eject
\section{Introduction}

It is a long lasting problem to determine  static
properties
of hadrons from the general theory of the strong interaction,
Quantum Chromo Dynamics (QCD) \quref{gross}. Therefore there were some
attempts in the past to derive an effective theory for the strong
interactions in some low energy approximation
\cite{bolo:cahill3,bolo:cahill,bolo:dype1,bolo:dype3,bolo:ball1}.
However
none of the derivations could exactly claim the range of validity of the
approximations. Although the resulting theory, which coincides
with the formerly  invented Nambu-Jona-Lasinio (NJL) model
\cite{bolo:njl1,bolo:njl2}, does not confine it shares
the maybe most important features of low-energy QCD relevant for ground
states.
These are: chiral symmetry and its spontaneous breaking.
In this case the lagrangian itself is chirally invariant but the
vacuum state breaks the symmetry. The symmetry breaking is driven
 by
a non-vanishing quark condensate, leading to the constituent masses of
the formerly massless  quarks. As a consequence
the Goldstone bosons emerge, namely  the pion, kaon and eta.
These almost massless excitations of ${\bar q}q$ pairs
               are expected to play a
dominant role in the QCD vacuum not only in the long range limit  but
also at small distances \cite{bolo:shuryak,bolo:shuryak2}.

In the presently investigated NJL model the nucleonic scenario is
realized by these Goldstone bosons and explicit quark degrees of
freedom. The gluon degrees of freedom are already implicitly contained
in the lowest order
because the action is obtained, at least formally,
after path integration  over the gluon
fields. The quarks are then bound in a selfconsistent potential
based on
a non-trivial chiral field configuration in the Hartree approximation
\cite{bolo:rewu,bolo:megrgo,bolo:wayo}
which is leading in $1/N_{\rm c}$ expansion. Though this picture
resembles much  an effective meson theory like the Skyrme model
\quref{skyrme}, one should note that
here the soliton is non-topological, i.e. practically speaking based on
the non-perturbative dynamics. On the contrary the Skyrme soliton is
topological, i.e. the baryon number totally hinges on the topological
winding of the chiral field, while in the NJL model the baryon number is
 carried by 3 (or \Ne) valence quarks.
On the basis of the analysis of the proton polarization
it was argued in Ref.\quref{forte1}
that the valence quark degrees of freedom,
missing in the Skyrme model, do account for some important
physics.

There is however another, more profound difference between the present
model and the effective meson theories. The effective meson theories are
based upon some lagrangian density being local in time and space,
whereas the NJL type
models are formulated in terms of the path integral over the fermions,
which can be regarded as a {\it time-ordered} product of the field
operators.  It has been  for a long time overlooked that this {\it
time-ordering} may bring up some new contributions for various baryonic
observables. In the recent papers \cite{bolo:wawa,bolo:ab9,bolo:allstars}
these contributions (hitherto referred to as {\it time-ordered})
have been calculated for axial decay constants and for
magnetic moments.

The example of the axial constants is perhaps the most persuasive.
It is
well known that in the nonrelativistic quark model
$g_{\rm A}=(\Nc+2)/3$. This means that there are important $1/\Nc$
corrections to $g_{\rm A}$, which for $\Nc=3$ amount to 60\% of the
leading result. In the effective meson theories the leading term for
$g_{\rm A}$ scales also as $\Nc$, however the next-to-leading
correction comes only at the $1/\Nc$ level in the SU(2) version of
the model. This is due to the fact that effective meson lagrangians are
{\it even} in field derivatives. In the cranking approximation for
the rotating soliton each time derivative counts as $1/\Nc$. The only
possible source for the contributions linear in the time derivative is the
Wess-Zumino term which vanishes identically in the SU(2) case. However in
the SU(3) case
both in the NJL and in the Skyrme model
such corrections appear. In the NJL model  in the leading order of
LWLA (Long Wave Length Approximation) or gradient expansion they are
equal to
the Witten's anomalous current \quref{wit83a} of a local mesonic theory.

It was always believed that the effective quark theories are equivalent
to the effective meson theories in a sense that they correspond to some
calculable local lagrangian density. Strictly speaking such a
lagrangian density can be derived from the NJL like model only in the
limit of the large soliton size. Although this statement remains true it
does not mean that the matrix elements of some operators, take as
example the axial currents, are the same irrespectively if they are
calculated straight away from the fermion path integral or from the
equivalent
local lagrangian density. This is due to the two facts: 1) upon the
semiclassical quantization of the soliton the cranking velocities are
promoted to the collective operators which do not commute with the
rotation matrix itself and 2) the path integral is {\it time-ordered},
{\it i.e.} it dictates unambiguously in which order the cranking
velocity and the rotation matrix appear in the expressions for the
matrix elements of the axial currents. If these  {\it non-local}
properties of the path integral are properly taken into account then
one gets the desired  $1/\Nc$ corrections. These corrections are
not small and improve the phenomenological predictions of the NJL
model. On the contrary in the {\it local limit} of the
present effective quark theory they are identically zero.

In the previous paper \quref{ab9} we have calculated the three SU(3)
axial decay
constants \gt, \go~and \gs~ in the chiral limit within the
semibosonized NJL model. This model
reproduces the hyperon spectra
\cite{bolo:ab4} and also the isospin splittings within baryon
multiplets \quref{ab5}.  Earlier the properties of the  axial currents
have been investigated in the
NJL model only for the case of SU(2) \cite{bolo:mego,bolo:wayo}.
Then the $1/\Nc$ corrections for SU(2) have been roughly estimated  in
\quref{wawa}, neglecting regularization and sea contribution and to full
extent in \quref{allstars}.  Beyond this there are
only calculations of the axial coupling constants within the
pseudoscalar SU(3) Skyrme model
\cite{bolo:psw} and the  pseudoscalar vector meson SU(3) Skyrme model
\quref{pw2}.
Actually hyperon spectra indicate \quref{ab4}
that the
scalar and pseudoscalar NJL model gives a more refined
structure
of the collective hamilton operator,
than the pseudoscalar Skyrme
model. A comparable structure can be obtained in the Skyrme model only
by introducing explicit vector mesons  and {\it modeling} the
anomalous and symmetry breaking part of the effective action.
This introduces a
large number of parameters,  whereas
in the NJL model  the parameters
\footnote{The free parameters are: $M$ -- constituent quark mass,
 and cut-off parameters and to some extent
$m_{\rm s}$ -- strange quark mass
}
can be fixed completely  by
requiring proper
mesonic masses and decay constants. The symmetry breaking pattern and
the anomalous effective action is then uniquely determined.

In contrast to our previous paper \quref{ab9}
now we implement explicit {\it time-ordering}
within the framework of the effective Euclidean action (EEA).
This treatment allows us to make a clear distinction
between the terms which emerge from the real or form the imaginary
(anomalous) part of the EEA.  Actually it turns out that
the new  explicitly {\it time-ordered} terms
emerge from the real part of the EEA and therefore have to
be regularized. In the present approach
the regularization prescription
is unique and the regularization function is derived in a well defined
manner.

In the present paper we furthermore extend our previous calculations and
calculate the $\ms$ corrections to the axial coupling constants.
Besides
the theoretical interest due to the new explicitly {\it
time-ordered} corrections,
the axial constants are of utmost phenomenological importance
as far as the
comparison with the recent  measurements
of the polarized proton and neutron structure functions is concerned.
So the main  phenomenological concern of this paper will be a comparison
of the model predictions with the experimental data for these
quantities. Especially we will concentrate on the role of the
explicitly {\it time-ordered} corrections and  furthermore
on the so called {\it anomalous} quantities, which are dominated by the
valence
contributions \cite{bolo:ab3,bolo:ab4,bolo:forte1}. The latter
manifest another conceptual difference between the Skyrme model and
the present NJL model.

The organization of the paper is as follows. In Sec. II we review the
basic features of the NJL model with special emphasis   on the solitonic
description. In Sec. III we describe the quantization procedure and
summarize the results on the hyperon splittings. We use
the mass splittings to
fix the parameters of the model. In Sec. IV
we derive expressions for axial currents in the chiral limit. Special
emphasis is put on the new contributions from the explicit
time-ordering and
their regularization. Then in Sect. V we discuss mass corrections to the
axial currents. Our numerical results are presented in Sect. VI.
Section VII contains a brief comparison with the results of the Skyrme
model. We present our conclusions in Sect. VIII.

In the Appendices we present
useful formulas for the semiclassical quantization (App. A),
a derivation of the regularization
functions (App. B),  a
gradient expansion for normal and time-ordered quantities (App. C).

\section{The SU(3) Nambu-Jona-Lasinio model - Solitons }
\label{sect2}

The quark Nambu-Jona-Lasinio model \cite{bolo:njl1,bolo:njl2}
can be written in the four-fermion formulation as:
 \beq {\cal
   L}_{\rm NJL} = {\bar q}(x)  ( i \dsl -  m ) q(x)  -2G \bigl[ ({\bar q}
   \lambda^a q)^2 + ({\bar q}i\gamma_5 \lambda^a q)^2 \bigr].  \label{g1}
		\eeq
Here the summation  over the $\lambda^a$ matrices is
implicit  (with $\lambda_0=\sqrt{2/3}$) and $m$ is the bare quark mass
matrix.  In the chiral limit  the Lagrangian has the desired
SU$(3)_{\rm R}\otimes$ SU$(3)_{\rm L}$ symmetry in addition to the
U$(1)_{\rm V}\otimes$ U$(1)_{\rm A}$, where the U$(1)_{\rm A}$  is the symmetry
which is not
shared by QCD. In principle one could introduce the 't~Hooft term into the
Lagrangian, which breaks the   U$(1)_{\rm A}$ explicitly. It could then serve
as a source for the $\eta'$ mass, which  otherwise  would be a
Goldstone boson.
This was recently done by {  Kato et al.} \quref{kato}
and the resulting profile for the SU(2) soliton was very similar to
the solutions that are restricted to the chiral circle (i.e. non-linear
case) and that are used here. So we conclude that
the effects of the 't~Hooft determinant on the solitonic observables of
the present calculations  are rather small.
However it was proven in \qref{kato} that the 't~Hooft term leads to a
stabilization of the  linear version of the model, which furthermore
supports the use of the non-linear Ansatz for the chiral fields.

Performing the bosonization procedure of
Eguchi \quref{egu} one arrives immediately   at the new  classical
Lagrangian
\footnote{
We will always  denote by E Euclidean and by M Minkowski quantities if
it is not obvious  from the context.}
\beq
  {\cal L}_{\rm E}  = {\bar q}(x) ( -i \dsl +
    g \left(\sigma^a \lambda^a + i \gamma_5 \pi^a \lambda^a \right)
    +  m ) q(x)  +
    {1\over 2}  \mu^2 \left( \sigma^a \sigma^a + \pi^a \pi^a \right),
     \label{g2}   \eeq
where the original coupling $G$ is given by $G=g^2/\mu^2$.
For convenience Eq.(\ref{g2})
can be rewritten in the polar decomposition of the chiral fields  \quref{wawe}:
\beq {\cal
   L}_{\rm E}  = {\bar q}(x) \left( -i \dsl + P_{\rm R}  {\cal M}  + P_{\rm
   L} {\cal M}^+ +  m \right) q(x)  + {1\over 4}  {\mu^2\over g^2}
	{\rm Tr}_\lambda \left( M M^\dagger \right), \label{g2a}   \eeq where
$ {\cal M} =  g (\sigma^a \lambda^a + i \gamma_5 \pi^a \lambda^a ) =
\xi_{\rm L}^\dagger M \xi_{\rm R}$ is given in terms of the unitary
matrices $\xi_{\rm L}$, $\xi_{\rm R}$ and a hermitian matrix $M$.  The
$P_{\rm R/L}=1/2(1\pm\gamma_5)$ are the right/left helicity projection
operators.  Furthermore  the unitary gauge $\xi_{\rm R}=\xi_{\rm
L}^\dagger=\xi$ can be chosen to eliminate the redundant degrees of
freedom.  As an approximation the scalar degrees of freedom in $M$ are
frozen, such that $M$ is just the constituent quark mass matrix and
$\xi$ is related to the
chiral field by $U=\xi^2$. The latter  can be parametrized
 as $\exp{(i \pi^a
\lambda^a/f_\pi)}$. This parametrization corresponds in SU(2) to the
chiral circle condition
$\sigma_{(2)}^2+{\vec\pi}^2={\rm const}$, where $\sigma_{(2)}$ is the  SU(2)
isoscalar $\sigma$ field.   Then  the
last term in
\qeq{g2a} is just a constant and will be omitted in the following.

Now we will shortly summarize  how to calculate quantum corrections to
${\cal L}_{\rm E}$ which are due to the fluctuations of the fermion fields.
The fermion determinant obtained by the integration over the quark
fluctuations
corresponds to the leading contribution in  $1/\Nc$ expansion.
The next to leading terms, the so
called boson determinant, have been shown to have minor influence on the
mesonic observables \quref{ab7}. Hence the effective Euclidean action
(EEA)
can be written as:
\beq  S_{\rm eff}[U(x)] =
     - \Sp \log \left(
     -i\dsl +  P_{\rm R}  {\cal M}(x)  + P_{\rm L} {\cal M}^\dagger(x)  +  m
\right),
    \label{g2b}
\eeq
where $\Sp$ is the functional trace in color, flavor and
momentum space.

 From
\qeq{g2b} the parameters
               of the model can be  fixed by requiring  experimental
values for the pion decay constant $f_\pi=93$~MeV, the pion and the kaon
mass,
$m_\pi=139$~MeV and $m_{\rm K}=496$~MeV,
(see \qref{ab4} for details). As a result the constituent quark mass is
the only free parameter of the model, which can e.g. be used to fix
baryonic properties \cite{bolo:ab4,bolo:ab5}.

Solitonic solutions of \qeq{g2b} can be found by making a
time-independent {\it hedgehog} Ansatz for the chiral field $U$ and writing
the SU(2) action in the chiral limit in terms of a single particle
intrinsic hamiltonian:
\beq
H=-i\gamma_4 \bigl(-i\gamma_i\partial_i + M U(x)\bigr).
 \label{1pH}
\eeq
We will specify the SU(3) extension of the SU(2) Ansatz in the next Section.
In the proper time regularization \quref{schwinger}, the effective
action becomes in this case:
\beq
  S_{\rm eff}  = \Sp \int {d u\over u } \phi(u)
  \exp{\left(-u(-\partial_\tau^2 + H^2)\right)},   \label{seff}  \eeq
where $\phi(u)=c\theta(u-1/\Lambda_1^2)+(1-c)\theta(u-1/\Lambda_2^2)$
is the regularization function of \qref{ab4}, which reproduces
common  values for the current quark masses and quark condensates in
the vacuum.
The classical equations of motion can be solved selfconsistently for the
chiral field $U$, resulting in a  localized soliton with unit winding
number. The  energy spectrum of the Hamilton operator $H$
(\qeq{1pH})
for the baryon number one sector
contains a discrete valence level inside
a mass gap of the size $2M$
\cite{bolo:megrgo,bolo:rewu,bolo:dipepo}.
Then the classical energy of the soliton can be written as
\cite{bolo:dipepo}:
\beq M_{\rm cl} = \Nc E_{\rm val} + \Nc E_{\rm sea},  \eeq
where $E_{\rm val}$ is the energy of the valence level and $E_{\rm sea}$
resembles the polarization the Dirac sea as a sum over the whole
spectrum of the Hamilton operator $H$.

\section{Quantization of Zero Modes and Mass Splittings}

The purpose of this section is to apply the semiclassical quantization
method to the solitons  \cite{bolo:megrgo,bolo:rewu,bolo:wa}  of Sect.
\ref{sect2}, which result from the classical and time-independent
equations of motion. The idea hereby is the
following. In order to quantize the system one can perform a
time-dependent transformation \quref{anw}, which can be in the
direction of the
symmetry or orthogonal to it. If the symmetry is at least an approximate
symmetry
then excitations in this direction should be the dominant contribution
to the low lying resonances of the model. In order to check this
numerically in the present model, we consider in addition to the usual
expansion of the EEA up to the second order in the rotational velocity
{\it consistently} the quadratic corrections from the strange symmetry
breaking terms.

Therefore following the treatment of
\qref{anw} we quantize the soliton by introducing time-dependent
SU(3) rotations and impose canonical quantization conditions for the
collective coordinates of the rotation matrix. This will allow for the
definition of generators of SU(3) and the corresponding baryon states.

First we make use of the  trivial embedding of   {  Witten}
\quref{wit83b} of the SU(2) chiral field  $U_0(x)=(\sigma_{(2)} +
i\gamma_5{\vec\tau}{\vec\pi})/f_\pi$ into the isospin subgroup of SU(3)
according to
\beq   U(x) = \left( \matrix{ &U_0 &0 \cr &0  &1\cr} \right) .
\label{g3}
\eeq
The
soliton solutions of SU(2) are also solutions for SU(3). In the
quantization procedure the embedding (\ref{g3}) generates,
as will be seen later explicitly, the correct
quantum numbers for baryons \quref{wit83b}.

Next one introduces a time-dependent rotation:
$ U(x,t)=A(t)\; U(x)\;A(t)^\dagger $. This rotation can be undone by
rotating the quark fields:
${\tilde q}=A(t)^\dagger q$ and  ${\tilde{\bar q}}={\bar q}
A(t)$. Then we obtain:
\bea
A^\dagger \dot{A} &=&
 i\Omega_{\rm E} ~~=~~\frac{i}{2}\lambda_a \Omega_{\rm E}^a
 \eea
and the relation between Euclidean and Minkowski velocities holds:
$i\Omega_{\rm E}=\Omega_{\rm M}$ and $\Omega_{\rm E}^\dagger=\Omega_{\rm E}$.

Expanding $S_{\rm eff}^{\rm rot}=-\Sp\log
(\partial_\tau+H+i\Omega_{\rm E}-i\gamma_4A^\dagger mA)$ of  \qref{ab4} up to
the quadratic order in
$\Omega$ (in Minkowski metric and in the chiral limit) one gets:
\beq
L_0 =-M_{\rm cl}+\frac{1}{2} \Omega_{a}I_{ab}\Omega_{b}
 - \frac{N_{\rm c}}{2 \sqrt{3}} \Omega_{8}
\label{eq:L}
\eeq
where tensor of inertia
$I_{ab}={\rm diag}(I_1,I_1,I_1,I_2,I_2,I_2,I_2,0)$
 can be found in \qref{ab4}.

The original path-integral $\int {\cal D}U(x,t)$ will
be in the following approximated by the integral  over the rotation
matrices $A(t)$ only, neglecting translations and other fluctuations
\quref{dipepo}.  This is known as the quantization of the rotational zero
modes \quref{anw}.  Instead of functionally integrating over all $A(t)$,
one makes usually use of the fundamental relation between the path-integral
and the hamiltonian operator formulation \quref{tdlee}:
\beq
        \int {\cal D} A(t) \exp{(-\int_{-T/2}^{T/2} dt \;
L_0^{\rm E})
}
         = < A(T/2) \mid \exp{(-T H^{(0)})}
         \mid A(-T/2) >,  \label{qmech}
\eeq
where
$H^{(0)}$ is a collective rotational hamiltonian corresponding to the
Lagrangian of eq.(\ref{eq:L}). In this way the path-integral can be
evaluated in terms of the eigenstates of the collective hamiltonian.
When we consider expectation values of currents similar relations hold.
There however one has to pay attention to the fact that the path-integral
is {\it time-ordered} in a natural way
\cite{bolo:tdlee,bolo:ramond},
whereas in the operator formalism
{\it time-ordering} has to be introduced explicitly.
The latter fact will be of importance when we consider
expectation values of currents  in the next section
and is  extensively described in the case of SU(2) in
\qref{allstars}.
Before we calculate axial properties we shall concentrate on mass
splittings. This allows to judge the perturbation expansion in $m_s$
and to fix the remaining parameter of the model - the constituent quark
mass M.

To calculate the  mass splittings one has to expand the effective
action
 in powers of the current quark mass $m=
\mu_{0}\,\lambda_{0} - \mu_{8}\,\lambda_{8} -\mu_{3}\,\lambda_{3}
$
with
\beq
 \mu_{0}  =  \frac{1}{\sqrt{6}}(m_{\rm u}+m_{\rm d}+m_{\rm s}),
{}~~\mu_{8}  = \frac{1}{\sqrt{12}}(2\,m_{\rm s}-m_{\rm u}-m_{\rm d}),
{}~~\mu_{3}  =  \frac{1}{2}\Delta m
\eeq
where $\Delta m=m_{\rm d}-m_{\rm u}$.

Here an important remark is in order. There are apparently two small
parameters in the present approach: 1/\Ne~ and $\ms$. Unfortunately from
the explicit calculations one cannot deduce  what actually sets the scale
for the $\ms$ corrections.
In any case  1/\Ne~ and $\ms$ can be
treated as being of the same order. Therefore to be consistent we
expand the effective action up to
terms of the order of $\ms$, $\ms^2$, $\ms \Omega$ and $\Omega^2$
 (expansion in
$\Omega$ corresponds to expansion in 1/\Ne):
\begin{eqnarray}
L_{m}~ & = &-\sigma m_{\rm s} + \sigma  m_{\rm s} D_{88}^{(8)},  \label{lm0m1}
\\
 L_{m \Omega} & = & -\frac{2}{\sqrt{3}} m_{\rm s} D_{8a}^{(8)}
K_{ab}{\Omega}_{b},  \label{lm0m2}  \\
L_{m^2} & = & \frac{2}{9} m_{\rm s}^2
(N_{0} (1-D_{88}^{(8)})^2+ 3 N_{ab} D_{8a}^{(8)} D_{8b}^{(8)}),
\label{lm0m3}
\end{eqnarray}
where the constant $\sigma$
is related to the sigma term
$\Sigma=3/2\;(m_{\rm u}+m_{\rm d})\;\sigma$ and
$D_{ab}^{(8)}=1/2\;{\rm Tr}(A^{\dagger}\lambda_a A \lambda_b)$.
Therefore we can define in this order $L^{tot}=L_0+L_m+ L_{m \Omega}
+L_{m^2}$.
The mass spectrum obtained with the help of $L_0+L_m+ L_{m \Omega}$
was discussed in
Refs.\cite{bolo:ab4,bolo:ab5}; there one can also find explicit
formulae for
$K_{ab}={\rm diag}(K_1,K_1,K_1,K_2,K_2,K_2,K_2,0)$.
Let us here only remind that the  {\it anomalous} moments of inertia $K_i$
are nearly entirely given by the valence part, whereas the contribution
of the valence level to $I_i$ amounts to approximately 60\%.
The new
feature of the present calculation is the presence of the
moments
of inertia $N_{ab}={\rm diag}(N_1,N_1,N_1,N_2,N_2,N_2,N_2,N_0/3)$
in  $L_{m^2}$
defined as:
\beq  N_{ab} = {N_c \over 4} \sum_{n,m}   <m\mid \lambda_a \gamma_0
\mid n> <n \mid \lambda_b \gamma_0 \mid m>
   {\cal R}_\beta (E_n,E_m),  \label{matn}
\eeq
where  ${\cal R}_\beta (E_n,E_m)$ is given by
\beq  {\cal R}_\beta (E_n,E_m)  =  { 1 \over 2 \pi }
\int { d u \over \sqrt{u} } \phi (u) \left[ { E_n e^{-uE_n^2} -
    E_m e^{-uE_m^2}  \over E_n - E_m }  \right],
    \label{regn}   \eeq
which differs from the regularization function for the usual moment of
inertia ${\cal R}_I (E_n,E_m)$ \quref{ab4} because of  the different
hermiticity behavior of the mass term and the Coriolis term  $(\Omega)$
in  $S_{\rm eff}^{\rm rot}$.

The values of $N_{0,1,2}$ together with the values of $I_{1,2}$ and
$K_{1,2}$ for different constituent masses are listed in \qtab{tabIKN}.

The Lagrangian of eq. (\ref{eq:L}) and eqs. (\ref{lm0m1}-\ref{lm0m2})
reminds the Skyrmion Lagrangian with vector mesons (c.f.  \qref{pw2}).   The
quantization
proceeds as in the Skyrme model; one defines the quantities
(see App. A for details):
\beq
J_a=-R_a=I_{ab}\Omega_b-\mu_i
D_{ib}K_{ba}-\delta_{a8}\frac{\Nc}{2\sqrt{3}}
\label{eq:Ja}
\eeq
($i=3$~and~$8$, $a,b=1\cdots 8$)
which are promoted to the spin operators ${\hat J}_a=-{\hat R}_a$.
The flavor operators read: $\hat T_a=-D_{ab} \hat J_{b}$. Note that
despite
the fact that $\hat J_a$ fulfil the SU(3) algebra, only
$\hat J_{1,2,3}$ have the
meaning of the symmetry generators. That is due to the structure of the
SU(3) {\it hedgehog} Ansatz and is reflected in the
fact that  $\hat J_8=-\Nc/\sqrt{12}$  generates a constraint.
Therefore the wave function of the baryon state $B=Y,T,T_3,J,J_3$
belonging to the SU(3) representation ${\cal R}$ reads (see App. A):
\begin{eqnarray}
\mid {\cal R},B >
 & = &\sqrt{{\rm dim}{\cal R}}
\left< Y,I,I_{3} \mid D^{({\cal R})}(A) \mid -Y^{\prime} ,J,-J_{3}
\right>^{\bf *},
\label{eq:wf}
\end{eqnarray}
where the right hypercharge $Y^{\prime}$ is in fact constrained to be
$ -1$.
The lowest SU(3) representations which contain
states with $Y=1$ are: ${\cal R}=${\bf 8} and ${\cal R}=${\bf 10}.
The quantized collective hamiltonian $H^{tot}$
 from
\beq   H^{tot} = \sum_a \Omega_a { \partial L^{tot} \over
         \partial \Omega_a}
                  -   L^{tot}      \label{legend}
\eeq
reads:
\beq
H^{(0)} = M_{\rm cl} + H_{\rm SU(2)} +H_{\rm SU(3)} ,
\label{ham0}
\eeq
\begin{eqnarray}
H_{\rm SU(2)}=     \frac{1}{2I_{1}}
C_{2}( {\rm SU(2) }, &~~~&
H_{\rm SU(3)}=
\frac{1}{2I_{2}}  \left[
C_{2} ({\rm SU(3)})-C_{2}({\rm SU(2)})-\frac{N_{\rm c}^{2}}{12}
\right].   \nonumber
\end{eqnarray}
Here $C_2$ denote the Casimir operators of the spin SU(2) and flavor
SU(3). $M_{\rm cl}$ is the classical soliton mass. It has been
calculated by many authors and its value turns out to be relatively large:
$M_{\rm cl}\approx 1.2$~GeV. This is a common problem for all chiral models.
 There are however some negative corrections to it, like Casimir energy
or rotational band corrections which
might bring $M_{\rm cl}$ to the right value.
In this paper, instead on
insisting on the calculation of the absolute masses, we will
concentrate on the mass splittings.
These are determined in the present model by analytic strange mass
contributions  in ${\cal O}(m_s)$ and ${\cal O}(m_s^2)$. Whereas linear
terms are given by eqs. (\ref{lm0m1},\ref{lm0m2}), the
the various contributions from quadratic $m_s$ corrections can be classified
into the following three cases:
\bit
\item quadratic terms from the expansion of the EEA,
corresponding to the Lagrangian in  eq. (\ref{lm0m3}).
This will be referred to as {\it kinematical} correction.
\item quadratic terms from replacing the rotational velocities in
\qeq{eq:Ja} by the generators $J_a$ by using \qeq{legend}.
This will be referred to as {\it dynamical} correction.
\item quadratic corrections from sandwiching the linear part
$L_m,L_{m\Omega}$, which can mix various representations of SU(3),
with linear mass corrections from the collective wave-function, as will
be discussed in the following.
This will be referred to as {\it wave-function} correction.
\eit

Then the hamiltonian from \qeq{legend} up to terms quadratic in $m_{\rm
s}$ reads:
\bea
H^{(1)}&  = & \left\{ \sigma
       - r_2 Y
       - (\sigma-r_2) D_{88} +
      \frac{2}{\sqrt{3}} (r_1-r_2)
      \sum\limits_{A=1}^3 D_{8A} J_A  \right\} \ms, \nonumber \\
H^{(2)}_{\rm kin} &  = & \frac{2}{3} \left\{
         r_2 K_2  ( 1 - D_{88}^2 )
       + (r_1 K_1-r_2 K_2)
       \sum\limits_{A=1}^3 D_{8A}^2  \right\} \ms^2, \label{eq:dynkin} \\
H^{(2)}_{\rm dyn} & = & -\frac{2}{9} \left\{ (N_0+3 N_2)
           -2 N_0 D_{88}
       +  (N_0- 3 N_2) D_{88}^2
       +  3 (N_1-N_2) \sum\limits_{A=1}^3 D_{8A}^2
       \right\} \ms^2, \nonumber
\eea
where $r_i=K_i/I_i$. According to items above we have split the
O($\ms^2$) hamiltonian into
the {\it kinematical} part $H^{(2)}_{\rm kin}$,
and  the {\it
dynamical} part $H^{(2)}_{\rm dyn}$.

The Hamiltonian $H^{(1)}$ mixes states of different SU(3)
representations,
therefore the wave function is no longer a pure octet or decuplet but
rather a mixture:
\bea
\mid  B~ > &=& \mid 8,B~> + c_{\overline{10}}^B~ \mid \overline{10},B~>
                       +c_{27}^B~ \mid 27,B~>, \nonumber \\
\mid B^{\prime} > &=& \mid 10,B^{\prime}> + c_{27}^{B^{\prime}}
\mid 27,B^{\prime}>
     +c_{35}^{B^{\prime}} \mid 35,B^{\prime}>  \label{eq:wf1}
\eea
where $B$=N,$\Lambda$,$\Sigma$,$\Xi$ and
$B^{\prime}$=$\Delta,\Sigma^*,\Xi^*,\Omega$. The coefficients $c_{\cal R}^B$
depend linearly on $\ms$, therefore with this accuracy there is no need
to change the normalization of the wave function.
In the following we will need their explicit form
only for the octet-like states:
\bea
c_{\overline{10}}^B = \frac{\sqrt{5}}{15} (\sigma - r_1)
\left[ \begin{array}{c}
 1 \\
 0 \\
 1 \\
 0
 \end{array}\right] I_2\; \ms, &~~~ &
 c_{         {27}}^B = \frac{1}{75} (3 \sigma + r_1 -4 r_2)
\left[ \begin{array}{c}
 \sqrt{6} \\
 3 \\
 2 \\
\sqrt{6}
 \end{array}\right] I_2\; \ms
 \eea
 in the basis $[{\rm N},\Lambda,\Sigma,\Xi]$. In Fig.1 we plot $c_{\cal R}^B$
in dependence on the constituent mass $M$.
The corresponding
O($\ms^2$) contribution to the energy reads:
\bea
E^{(2)}_{\rm wf}&=&- \left\{\frac{1}{60}
\left(1+ Y -X+ \frac{1}{2} Y^2 \right)
(\sigma-r_1)^2
\right.
\nonumber \\
 & &~~~\left.
 +\frac{1}{250} \left( \frac{13}{2}+\frac{5}{2}X -\frac{7}{4} Y^2 \right)
 \frac{1}{9}(3 \sigma + r_1 - 4 r_2)^2
 \right\} I_2 \ms^2 \label{eq:wf8}
 \eea
 for the octet and for the decuplet:
\bea
 E^{(2)}_{\rm wf} &=&- \left\{ \frac{1}{16}
 \left(1+ \frac{3}{4}Y+ \frac{1}{8} Y^2 \right)
 \frac{1}{9}(3 \sigma - 5 r_1 + 2 r_2)^2
 \right.
 \nonumber \\
 & &~~~\left.
 +\frac{5}{336} \left( 1 - \frac{1}{4} Y - \frac{1}{8} Y^2 \right)
(\sigma + r_1 -2 r_2)^2
\right\} I_2 \ms^2. \label{eq:wf10}
 \eea
Here
$X=1-I(I+1)+1/4\;Y^2$ is the usual combination entering
Gell-Mann--Okubo mass relations.

With the help of the  matrix elements of the $D$ functions and spin
operators discussed in the App. A one arrives at the following
result for the hyperon splittings:
\bea
\Delta M^{(8)} &=& A -\frac{F}{2}\;Y-\frac{D}{\sqrt{5}}\;X - G\;Y^2,
\nonumber \\
\Delta M^{(10)} &=& B -\frac{C}{2\sqrt{2}}\;Y- H\;Y^2.
\label{abcdfgh}
\eea
Constants $A$ and $B$ do not contribute to the splittings within the
multiplets, however they shift the mass centers and contribute to the
{\bf 10}-{\bf 8} mass difference. Constants $G$ and $H$ not present in the
first
order Gell-Mann--Okubo mass formula are of the order of $\ms^2$.
Experimentally one gets:
\bea
F &=&\Xi-{\rm N}=379~{\rm MeV},\nonumber \\
D & =&\frac{\sqrt{5}}{2}(\Sigma-\Lambda)=86~{\rm MeV},\nonumber \\
G&=&\frac{1}{4}(3\Lambda+\Sigma)-\frac{1}{2}({\rm N}+\Xi)=6.75~{\rm MeV}
\label{eq:8exp}
\eea
for the octet. For the decuplet the three operators: 1, $Y$ and $Y^2$ do
not form a complete basis and therefore there are two independent
relations which determine constants $C$ and $H$ with some small uncertainty:
\bea
C=\sqrt{2}(\Xi^* - \Delta)&=&\frac{1}{\sqrt{2}}(\Omega-2 \Delta+\Sigma^*)=
422.5\pm3.5~{\rm MeV}, \nonumber \\
H=\frac{1}{2}(2\Sigma^*-\Xi^*-\Delta) & = &
\frac{1}{6}(3\Sigma^*-2\Delta-\Omega)=2.83\pm 0.33~{\rm MeV}.
\label{eq:10exp}
\eea

In \qtab{tababcdfgh}
we list the coefficients $A\ldots H$ for a typical value of
$\ms=180$~MeV
as functions of the constituent mass $M$. It can be seen
that in order to reproduce the experimental numbers of
eqs.(\ref{eq:8exp},\ref{eq:10exp}) one has to take
the constituent mass of the
order of 400~MeV. Then all constants $A\ldots H$ are roughly reproduced.
The constant $G$ and $H$ being of the order $O(\ms^2)$ are small.
For reasonable
strange quark  masses $O(\ms^2)$ corrections
to $A,B,C$ and $D$
are of the order of 20\%
of the leading $O(\ms)$ terms with the exception of $F$ for which
$O(\ms^2)$ corrections are almost zero.

In order to make phenomenological statements
we adopt the following procedure: first for given
$M$ we find the optimal $\ms$  which reproduces {\bf 10}-{\bf 8} splitting.
To this end we define the mean octet and decuplet values:
${\overline M}_8 = 1/2\;(\Lambda + \Sigma)=1155$~MeV and
${\overline M}_{10} = \Sigma^*=1385$~MeV. Then
$\Delta_{10-8}\equiv {\overline M}_{10} -{\overline M}_8 =230$~MeV
is given by:
\beq
\Delta_{10-8}=\frac{3}{2 I_1} + B - A. \label{eq:10-8}
\eeq
Since $A-B={\rm const.}\times \ms^2$ one can numerically solve
eq.(\ref{eq:10-8}) for $\ms$. The result is plotted in Fig.2.

In Fig.3 we show the $\ms$ dependence of the deviations
{\it theory -- experiment} for each hyperon.
On should remember that for each $\ms$
the optimal constituent quark mass $M$ was used, so that
$\Delta_{10-8}$ was automatically reproduced for each $\ms$.
 The smallest deviations $\pm 7$~MeV for all splittings correspond to
$\ms\approx 185$~MeV, {\it i.e.} $M\approx~425$~MeV.

It is interesting to examine to what extent the new corrections
calculated in this paper are important. The
Yabu--Ando method of diagonalizing the hamiltonian of $O(m_s)$
exactly is widely spread in the literature
\cite{bolo:ya,bolo:pw2,bolo:ab4,bolo:wealre}
and it essentially corresponds, in our language, to taking into account
only the wave function corrections of eqs.(\ref{eq:wf8},\ref{eq:wf10}).
Indeed for $\ms$ of the order of 200~MeV the second order wave function
correction almost exactly coincides with the exact result of the Yabu--Ando
method. This is illustrated in Fig.4. However consistency requires to
include the {\it kinematical} and {\it dynamical} contributions of
eqs.(\ref{eq:dynkin}) in the same order of $\ms$. Whereas the
{\it kinematical} corrections are always small the {\it dynamical}
ones are by no means negligible. This is explicitly shown in
\qtab{tabABCDFGH}, where all contributions to constants $A\ldots H$
are displayed for $M=423$~MeV and $\ms$=180~MeV.
Note that  $O(\ms^2)$ corrections to  $F$
are negligible.

The message here is clear: The wave function
contributions (w.f.) are not the whole story and one has to include
consistently all corrections, i.e. in addition kinematical and
dynamical contributions, in a given order of
$\ms$.

The purpose of this section was twofold: First we have demonstrated the
importance of the $O(\ms^2)$ corrections coming from the effective action as
compared to the wave function corrections
(the Yabu-Ando approach). Second
we have used the mass splittings to fix the parameter of the model, namely
the constituent mass $M$.  Having done this we can proceed to the
evaluation of the axial coupling constants.

\section{
Axial Currents in Chiral Limit }

In order to calculate observables like the  axial vector currents,
one has to consider the path integral expectation value of these
operators. This can be also done within the formalism of
the quark correlation
functions, see  e.g.
Refs.\cite{bolo:wayo,bolo:dipepo}.  Here will use however the
effective action approach
and show how the {\it time-ordering} within a quark loop together with
the collective quantization brings  up the corrections linear
in the rotational velocity $\Omega$
\cite{bolo:wawa,bolo:ab9,bolo:allstars}.
The approach presented here
will be different from the one of Ref. \cite{bolo:ab9}, where these new
corrections were calculated from the unregularized expressions.

One can express the axial vector current
$A_\mu^a$ as a path integral expectation value
\bea
     &   <A_\mu^a(x)>&=    N \int {\cal D}{\bar q}
        {\cal D} q  \int   {\cal D} U
       \bigl(  {\bar q} \gamma_\mu\gamma_5\lambda^a q \bigr)
    \; e^{-\int d^4x {\cal L}_{\rm E}  }  \nn
      &     &= {\delta\over\delta s(x) }{ \int {\cal D}{\bar q}
          {\cal D} q  \int   {\cal D} U
     \; {\exp \left\{ {-\int d^4x \left( {\cal L}_{\rm E}
-s\; {\bar q}
\gamma_\mu\gamma_5 I_a
       q \right)  }\right\}}
        \linie}_{s=0}
       \label{g4}
\eea
with the convention $\gamma_0=-i\gamma_4$  and the following
definitions
for $I_a$:
\beq g_{\rm A}^{(0)} : I={\bf 1}, \ \
     g_{\rm A}^{(3)} : I=\lambda_3,  \ \
     g_{\rm A}^{(8)} : I=\lambda_8.    \eeq
As in Sect. III
the quantization is performed by introducing the time-dependent
SU(3) rotation matrix $A(t)$.
Then we obtain
\beq
        <A_\mu^a(x)>= {\delta\over\delta s(x) }{ \int {\cal
         D} {\tilde {\bar q} }
          {\cal D}{\tilde q}  \int   {\cal D} U \;
    { \exp \left\{ {-\int d^4x \left( {\tilde{\cal L}}_{\rm E}
-s\; {\tilde{\bar q}}
    \gamma_\mu\gamma_5  A^\dagger \lambda^a A
     {\tilde q} \right)  }\right\} }} \linie_{s=0}   \label{g5}
\eeq
with
${\tilde q}=A(t)^\dagger q$ and  ${\tilde{\bar q}}={\bar q} A(t) $,
and the
  rotated lagrangian  ${\tilde{\cal L}}_{\rm E} $
\beq {\tilde{\cal L}}_{\rm E}  =    {\tilde{\bar q}} \left(
    -i\dsl +M\ U(x) +A^\dagger m A -i\gamma_4 A^\dagger {\dot A}
         \right)    {\tilde q}  .  \eeq
Integrating over the quark fields  and restricting the $ {\cal D} U$
integration to the SU(3) rotations \cite{bolo:dipepo}
gives for the space components of the current
\beq <A_i^a(x)>=
{\delta\over\delta s(x) } { \int {\cal D} A(t) \; \Spto \log D[s]
}\linie_{s=0}   \label{g5b}, \eeq
where $D[s]= \partial_t + H +  A^\dagger {\dot A}
  -i\gamma_4  A^\dagger m A
+i \ s\;\gamma_4\gamma_i\gamma_5  A^\dagger \lambda^a A$.

Expression (\ref{g5b}) needs now a careful explanation. This is due to the
fact that we are not going to perform the path-integral over the rotational
matrices but, instead, we will use the operator formalism after the
quantization of the generalized SU(3) coordinates has been
performed.  Within the path integral the {\it time-ordering} of the
operators after  the ${\cal D}A$ integration
is given in a natural way (see e.g.
\qref{ramond}).  In the operator formalism we can make use of the trace
properties only if we respect the {\it time-ordering} of the operators
(intrinsic and collective) within the
trace.  This is denoted in short by the modified trace $\Spto$, which
has the usual properties except for the time component. Only {\it after}
the explicit time ordering of the operators the rotational frequencies
$\Omega$ can be again considered
as {\it time-independent}.

One great advantage of retaining the trace in this form is
the straightforward applicability of the regularization procedure.
Let us split the effective action into real and imaginary part:
\bea
     <A_i^a(x)> &=&
     \int {\cal D} A(t)
     {\delta\over\delta s(x) } {  \left[
     {\rm Re}\; \Spto \log  D +
  i \; {\rm Im}\; \Spto \log  D \right]} \linie_{s=0}
\nonumber	\\
&=& \int {\cal D} A(t) {\delta\over\delta s(x) } \half  { \left[
 \Spto    \log  D^\dagger D   +
     \Spto   \log {D\over  D^\dagger}    \right]
      \linie}_{s=0}.       \label{g5c}    \eea
It is important to consider $s(x)$  as explicitly time-dependent (see
App. B).
	Preserving  vector gauge invariance \quref{ball2}  by
using the proper time regularization \quref{schwinger}
we regularize the real part of the effective action:
\beq  \Spto    \log  D^\dagger D
      \rightarrow  \Spto  \int {d u    \over u  } \phi (u)
      \exp{\left( - u   D^\dagger D\right)},  \label{g5d}   \eeq
where $\phi(u)$ is given in Sect. II.
Note that for {\it symmetric} contributions \quref{ab9}, i.e. when
the index of a generator $\hat{J}_a$ is such that it commutes
 with the $D_{bc}$-function, the
time-ordering has no influence.

In the following the separation into valence and sea part is done by
introducing  $D^\prime=D-\mu$, where $\mu$ is a chemical potential
with $0<\mu<E_{\rm val}$ \quref{megrgo}, such that
$S_{\rm eff}^{\rm tot}=S_{\rm eff}^{\rm val}+S_{\rm eff}^{\rm sea}$
with  the definitions $S_{\rm eff}^{\rm val}=S_{\rm eff}[D^\prime]
-S_{\rm eff}[D]$ and $S_{\rm eff}^{\rm sea}=S_{\rm eff}[D]$.
The subtraction of possible vacuum contributions is implicitly
understood.

\subsection{The lowest order contribution $\sim\Omega^0$}

The axial vector coupling constants $g_{\rm A}^a$, defined as the
corresponding formfactor in the limit $q^2=0$,
can be calculated from \qeq{g5c}.
Noting that the functional integral ${\cal~D}A$ over the rotation
matrices can be replaced by an ordinary integration $\int d\xi_A$ over
the collective coordinates one can define an operator
${\hat g}_{\rm A}^a$  such that
\beq   g_{\rm A}^a = \int d\xi_A\ <B(\xi_A)\mid \ {\hat g}_{\rm A}^a \
\mid B(\xi_A)>,
\label{g5d1} \eeq
where $\mid B(\xi_A) >$ is the baryon wave-function of
\qeq{eq:wf1}. Therefore ${\hat g}_{\rm A}^a$ is obtained by expanding
$<A_i^a(x)>$ in \qeq{g5c} in terms of the rotational
frequency and strange mass but without performing the
${\cal~D}A$-integration. The rotational velocity is then replaced
by the generators of \queq{eq:Ja} yielding after integration over
3-dimensional space  the collective operator
${\hat g}_{\rm A}^a$ in terms of $J_a$ and Wigner functions $D_{ab}$.

Therefore the lowest order result in
$\Omega$ (i.e. $\Omega^0$) comes from the proper-time
regularized real part of the EEA
(\ref{g5d}).
One obtains
for $a=3$ and 8 (see App. B for details):
\beq
    {\hat g}_{\rm A}^a (\Omega^0,m_s^0)
        =  M_3           D_{a3},    \label{g5e}
\eeq
where $A^\dagger I_a A = D_{ab}\lambda_b$ for $a=3$ and 8, and
$A^\dagger I_aA =1$ for $a=0$.  At this level ${\hat g}_{\rm A}^0\equiv
0$.
The quantity  $M_3=M_{3,{\rm val}}+M_{3,{\rm sea}}$ which comes from the
real part of the action is then given by:
\beq    M_{3,{\rm val}} = N_{\rm c}
< v\mid \gamma_0 \gamma_3\gamma_5 \lambda_3 \mid v>    \label{g5f}
\eeq
 and
\beq M_{3,{\rm sea}}   = -{N_{\rm c} \over 2} \sum_{all\ n} <n\mid \gamma_0
			 \gamma_3\gamma_5 \lambda_3 \mid n> \sign (E_n) {\cal
	    R}_\Sigma(E_n),     \label{g5g} \eeq
 where the regularization function reads \quref{ab3}:
\beq  {\cal R}_\Sigma(E_n) = {1\over
	 \sqrt{\pi} } \int_0^\infty {d u   \over \sqrt{ u  } } e^{- u  }
	 \phi({ u  \over E_n^2}).    \label{g5h}
\eeq
The values of $M_3$ are displayed in \qtab{tabM3M44}.
\subsection{Anomalous $1/\Nc$ corrections from the
imaginary part of the EEA}

Taking $1/\Nc$ corrections (i.e. terms linear in the rotational
velocity $\Omega$) into
account, one can make the clear separation into {\em local} quantities
which emerge from the imaginary part of the effective Euclidean action, and
{\it non-local} quantities
emerging from the real part due to explicit time-ordering of collective
operators.
It will
turn out that the former quantities are related to Witten's anomalous
axial current \quref{wit83b}, whereas the latter ones have no counterpart
in a mesonic effective theory.
In a certain
sense they renormalize the leading contribution of the axial current given
by eq.(\ref{g5e}) (compare with Refs.  \cite{bolo:dama1,bolo:dama2}).

In the chiral limit the anomalous corrections
linear in $\Omega$  can be written as:
\beq {\hat g}_{\rm A}^a
     = -  M_{bc} i \{ D_{ab}, \Omega_{\rm E}^c\}
     = -2 M_{bc}        D_{ab}\Omega_M^c   \label{g12}   \eeq
with
\beq   M_{bc} = {N_{\rm c}\over 4} \sum_{n,m}
           < n \mid \sigma_3 \lambda_b \mid m > < m \mid \lambda_c
         \mid n >
         {\cal R_M} (E_n,E_m)
       \label{g13}
\eeq
and the cutoff independent regularization function ${\cal R_M}$
 \beq
      {\cal R_M} (E_n,E_m) =
     \half  {   \sign (E_n-\mu) -  \sign (E_m-\mu)
           \over   E_n      -       E_m       }.
           \label{g14}
\eeq
As noted above the chemical potential $\mu$ is chosen in such a way, that it
always
lies between  the {\it valence level} and the positive continuum of
states.
In this way the quantities $M_{bc}$  correspond to the full baryon number one
contribution          and therefore contain the sum of the valence  and
the sea part.  Additionally we define for
later use
$\overline{ M}_{8a}=\sqrt{3}M_{8a}$ and $\overline{ M}_{a8}=\sqrt{3}M_{a8}$.
The only non-vanishing contributions in  \qeq{g12} are:
 $M_{83}$ and $M_{44}=M_{55}=-M_{66}=-M_{77}$.
Using the symmetries of the {\it hedgehog} states
one can write for
$a=3$ and 8:
\beq
      {\hat g}_{\rm A}^a  =
       -{2 \overline{ M}_{83}\over\sqrt{3}}   D_{a8} \Omega_3
       -4 M_{44}  d_{3bb}  D_{ab} \Omega_b,  \label{g15}
\eeq
where the sum over $b=4\ldots 7$ is understood.
The values of the coefficients entering eq.(\ref{g15}) are displayed in
Tabs. \ref{tabM3M44} and \ref{tabMR83}. It is clear from the form of
eq.(\ref{g15}) that the anomalous corrections linear in $\Omega$ vanish in
the SU(2) case.

\subsection{Normal $1/N_{\rm c}$ corrections
 from the real
part of the EEA}

Apart from the anomalous contributions of the preceding subsection,
which have their counterparts within the SU(3) Skyrme model
\cite{bolo:pw1,bolo:pw2}, there exist corrections linear in
$\Omega$, which come from the real, non-anomalous part of the EEA
and which are due to some explicit time-ordering of operators.
They were
recently discussed within the SU(2)
\cite{bolo:wawa,bolo:allstars}
and SU(3) \cite{bolo:ab9} version of
present model. As advertised in the beginning of this section
these terms emerge
because the operators $\hat{J}_a$ and $D_{ab}$ in the space of the
collective coordinates are in general time-dependent and have to be
time-ordered, if one makes use of the operator formalism \quref{dipepo}.
This is described at length in Appendix B  and  as a result the
contribution to   ${\hat g}_{\rm A}^a$  from these terms can be
summarized as:
\beq
    {\hat g}_{\rm A}^a  =
     {N_{\rm c}\over 4} i \left[ \Omega_{\rm E}^c,D_{ab}\right]
      \sum_{m,n} <n\mid\lambda^c\mid m><m\mid\sigma_i\lambda^b\mid n>
      {\cal R_Q} (E_n,E_m)
\label{g20}
\eeq
where the rather complicated  regularization function
${\cal R_Q}$ is given by:
\beq  {\cal R}_Q (E_n,E_m) =
      \int_0^1{ d\alpha  \over 2 \pi} { \alpha E_n-(1-\alpha) E_m \over
       \sqrt{\alpha (1-\alpha)}  } \; c_i \;
     { \exp{(-[\alpha E_n^2+(1-\alpha)E_m^2]/\Lambda_i^2)}
      \over
      \alpha E_n^2+(1-\alpha)E_m^2   }.    \label{aa24} \eeq
Here
the proper-time $u$-integration for our step-like functions
$\phi(u)$ has been already performed (see App.B for a general expression).
In the limit $\Lambda_i \rightarrow \infty$  \qeq{aa24}
immediately reduces to \qeq{aa26} of App. B and coincides therefore
with our former prescription in \qref{ab9}.  However, as we will see later,
with the regularization properly taken into account, the physical values
will come out much better.

Using the quantization condition for $\Omega$ and
making use of the commutator
$\left[\hat{J}_c,D_{ab}\right]=if_{cbd}D_{ad}$
\quref{toyota} \qeq{g20} can be written as:
\bea  {\hat g}_{\rm A}^a  &=&
           {-if_{cdb}D_{ad}\over I_{cc}}  Q_{bc}
\nonumber                        \\
     &=&   -\left( {2iQ_{12} \over I_1}
           +{2iQ_{45} \over I_2} \right) D_{a3},
      \label{g21}
\eea
where the  quantities $Q_{bc}$ coming from the real part of the EEA
are given by   $Q_{bc}=Q_{bc,{\rm val}}+Q_{bc,{\rm sea}}$. Explicitly
the valence part reads:
\beq  Q_{bc,{\rm val}} =  {N_{\rm c}\over 2} \sum_{n}
       {<n\mid   \sigma_3   \lambda_b \mid v>
        <v\mid  \lambda_c \mid n>   \over
           E_n-E_v  } \;  \sign E_n
\label{g50}  \eeq
and the sea part:
\beq  Q_{bc,{\rm sea}} =
         {N_{\rm c}\over 2} \sum_{m,n}
        <n\mid \gamma_0 \gamma_3\gamma_5 \lambda_b \mid m>
        <m\mid  \lambda_c \mid n>
                       {\cal R_Q} (E_n,E_m).
      \label{g51}  \eeq
The numerical values for the  $Q_{bc}$ can be found in \qtab{tabMpmM45}.

One should note however that the appearance of $Q_{bc}$ is due to the
fact that the operators in the EEA are explicitly time-ordered
and that in addition the matrix elements
$<n\mid\gamma_0\gamma_3\gamma_5\lambda_b\mid m><m\mid\lambda_c\mid n>$
are antisymmetric with respect to the interchange of $m$ and $n$.

The valence contribution
$Q_{bc,{\rm val}}$ differs from the formula given in \qref{wawa},
where the existence of such corrections was claimed
 for the first time.
The correct path-integral formula is
given by \qeq{g50}  and   \qeq{g51}.  Numerically however the difference
between our expression for  $Q_{bc,{\rm val}}$ and  the expression of
\qref{wawa} is quite small. Note also that in \qref{wawa}
 the sea contribution to $Q_{bc}$
was erroneously claimed to be identically zero. Again numerically
$Q_{bc,{\rm sea}}$ is rather small.

Putting all these corrections together one obtains:
\bea  {\hat g}_{\rm A}^a(\Omega^0+\Omega^1,m_s^0)  &=&
          \left[   M_3
           -{2iQ_{12} \over I_1}
           -{2iQ_{45} \over I_2} \right]   D_{a3}  \nn
    & &  -{2 \overline{ M}_{83}\over \sqrt{3} I_1} \;  D_{a8} \hat{J}_3
          -4 { M_{44}\over I_2}\; d_{3bb}  D_{ab} \hat{J}_b
      \label{g52}
\eea
($b$ runs over $4\ldots 7$).
Note that all the quantities  $M_3,Q_{bc},M_{bc}$
and also $I_{1,2}$ are of the order $O(N_{\rm c})$, such that the
$Q_{bc}$-terms  in the brackets  indeed correspond to $1/N_{\rm c}$
corrections to the lowest order result. In other words as far as one
neglects the anomalous, purely SU(3) contribution in \qeq{g52}, the ratio of
different ${\hat g}_{\rm A}^a$'s  has no $1/N_{\rm c}$ correction.

\subsection{The anomalous singlet axial current }

The singlet axial vector current was already given in
\quref{ab6}  and it gets only anomalous contribution linear in $\Omega$:
\beq
       {\hat g}_{\rm A}^0(\Omega^1,m_s^0)
       =-{2\overline{ M}_{83}\over I_1} {\hat J_3}.
    \label{g53}
\eeq
Note that \qeq{g53} given here in the context of SU(3) coincides
exactly with the SU(2) result of \qref{wayo}. This is because only spin
eigenvalues
$(J_3)$ enter here, whereas the other ${\hat g}_{\rm A}^a$'s
always contain $D$-functions, whose matrix elements  depend crucially on
the SU($N_{\rm flavor}$) algebra used.

\subsection{The axial currents in the leading order LWLA}

For large size of the soliton one can approximate the
different contributions in \qeq{g52} by the
gradient expansion (or long wave-length approximation (LWLA)) \quref{ait}.
The lowest order
result for SU(2) is given in \qref{mego} and
it coincides with the expressions
from the Skyrme model.
Terms linear in $\Omega$ can be also gradient-expanded.
In SU(3) one gets the anomalous
contribution coinciding with the Wess-Zumino-Witten
term in the Skyrme model. Using the
results of App.  C one can also calculate the LWLA of the {\it
time-ordered} terms\footnote{The moments $Q_{bc}$ are evaluated
here in the infinite cutoff limit in order to get a simple and cutoff
independent result.}.
Altogether one obtains:
 \bea {\hat g}_{\rm A}^a &=& \int dr  r^2 \left(
\theta^\prime +{\sin 2\theta\over r} \right) \left[ {8\pi\over 3} f_\pi^2
+{M \over 4I_1} + {M\over 8I_2} \right] D_{a3} \nonumber \\ & & - {4\over
I_2}  {N_{\rm c}\over 6\pi} \int dr r \; \theta^\prime \sin^3\theta(r) \ \
d_{3bb} D_{ab}{\hat J}_b. \label{g54} \eea
It is clear from \qeq{g54} that the {\it time-ordered}
contributions (i.e. $Q_{bc}$) lead to the renormalization of
$g_{\rm A}$ in the sense that they are also proportional to   $D_{a3}$.
In SU(2) where the second line of \qeq{g54} vanishes
the ratio of  $g_{\rm A}^a$'s for  different baryons  has no
$1/N_{\rm c}$ correction.  This was also found by Dashen and Manohar from
large $N_{\rm c}$ QCD \cite{bolo:dama1,bolo:dama2}.  Furthermore the SU(2)
result resembles very much the old non-relativistic quark model prediction
for the $1/N_{\rm c}$ correction which is given by: $g_{\rm A}=\Nc/3+2/3$.
Using the value $I_1=1.156~\fm$ for $M=423~\MeV$ from \qtab{tabIKN},
the SU(2) part of \qeq{g54} gets approximately $50\%$ correction from the
$1/N_{\rm c}$ term. In SU(3) however, there are additional corrections from
the third term in the brackets of \qeq{g54} and from the anomalous terms,
such that  the simple rescaling factor does not exist any more.
\section{Mass Corrections for ${g}_{\rm A}$}

In this Section we will evaluate the symmetry breaking corrections to
the axial currents due to the non-vanishing strange quark mass.
These arise from the term $A^\dagger mA=\mu_0-\mu_8\lambda^a\;D_{8a}$.
In the linear order in $m_s$ and in the zeroth order in $\Omega$
neither   contributions from the  imaginary part nor  from the explicit
time-ordering (because D-functions always commute with each other) exists.
Therefore entire symmetry breaking contribution comes from the real part of
the EEA. Performing the expansion of the real part of the EEA in $\ms$ one
gets:
 \bea {\hat g}_{\rm A}^a (\Omega^0,\ms^1)& =&
		  -{4\ms\over\sqrt{3}} R_{38} D_{a3} ( 1 - D_{88} )
          +{4\ms\over\sqrt{3}} R_{83} D_{a8} D_{83} \nonumber \\
 & &      +{8\ms\over\sqrt{3}} d_{3bb} R_{44} D_{ab} D_{8b}
\label{g16}
\eea
with $b=4\ldots 7$
 The proper time regularized quantities
		  $R_{bc}=R_{bc,{\rm val}}+R_{bc,{\rm sea}}$ are given by:
\beq    R_{bc,{\rm val}} = {N_c \over 2} \sum_n
       { < n \mid \sigma_3 \lambda_b \mid v > < v \mid \lambda_c
        \gamma_0
       \mid n > \over      E_n      -      E_v      }
\eeq
and
\beq   R_{bc,{\rm sea}} = {N_c\over 4} \sum_{n,m}
           < n \mid \sigma_3 \lambda_b \mid m > < m \mid \lambda_c
          \gamma_0
         \mid n >
         {\cal R}_\beta   (E_n,E_m).      \label{g17}
\eeq
with  \beq
      {\cal R}_\beta  (E_n,E_m) =
      {1 \over 2 \sqrt{\pi}} \int_0^\infty {dt\over\sqrt{t}} \phi(t)
      \bigl[ {E_n e^{-tE_n^2} - E_me^{-tE_m^2} \over
             E_n - E_m }           \bigr]
   \label{g18}
\eeq
For future use we also define
$\overline{ R}_{83}=\sqrt{3}{R}_{88} $  and
$\overline{ R}_{38}=\sqrt{3}{R}_{38}$.
Note that  ${\cal R}_\beta (E_n,E_m)$ is different from the
regularization functions ${\cal R_I} (E_n,E_m)$ and
${\cal R_Q}(E_n,E_m)$.
The origin of
this difference, which however survives only in the finite cutoff
case, is the different hermiticity behavior of the current and the mass
term on the one hand and the Coriolis term $i\Omega_{\rm E}$ in \qeq{g5b} on
the other hand. The latter one turns out to be antihermitian in
Euclidean space, whereas the former ones are hermitian.
Because the proper time regularization rests on building $D_{\rm E}^\dagger
D_{\rm E}$ from the very beginning, different signs emerge and lead to
different regularization functions. Their substantial different
behavior can be seen in Fig. 1 of \qref{ab6} and Fig.5 of the present
paper. We list the new coefficients
$R_{ab}$ in  \qtab{tabMR83} and \qtab{tabR44R38}.

Apart from   these terms from the action we have in addition the
$m_s$ terms from the quantization condition  \queq{eq:Ja}.
Including these mass corrections
and the ones from the expansion of the effective action we obtain
up to the linear order in the symmetry breaking and the rotational
frequency (for $a=3$ and 8):
\begin{eqnarray}
      \hat{g}_{\rm A}^a  &=&  \left[   M_3 -{2iQ_{12}\over I_1}
          -{2i Q_{45} \over I_2}  \right]
         D_{a3}
         - {4 M_{44} \over I_2} d_{3bb} D_{ab} \hat{J}_b    \nn
    & &  -  {2 \overline{M}_{83} \over \sqrt{3} I_1}   ( 1
           +{4\ms\over\sqrt{3}}
            {K_1\over I_1} D_{83} )\; D_{a8}  \hat{J}_3
 \nn
    & & +   { 4\ms\over \sqrt{3} } {R}_{83}  D_{a8}D_{83}
        + {8\ms\over \sqrt{3}} ( R_{44} - M_{44} {K_2\over I_2} )
       d_{3bb} D_{ab} D_{8b}     \nn
  & &  -{4 \ms\over \sqrt{3} } R_{38} D_{a3} ( 1-D_{88} ),
   \label{gai2}
\end{eqnarray}
where, as usually, the index $b$ in $d_{3bb}$ runs over $4\ldots 7$.

The quantities $R_{83}$  and   $M_{83}$
are already known from the expression of the flavor singlet
axial  constant \cite{bolo:ab6}.
We found there\footnote{Comparison with  Ref.\cite{bolo:ab6} can be
done by identifying
$\overline{ M}_{83}=\beta_1$ and $\overline{ R}_{83} =\beta_2$. Note also that
the  sign  in  Eq.(5)  in  Ref.\cite{bolo:ab6}  is  misprinted.
}  in the same order of the expansion:
\beq  \hat{g}_{\rm A}^0 = - {2 \overline{ M}_{83} \over I_1} \hat{J}_3
		  - {4 \ms \over \sqrt{3}}
     D_{83} \left( {K_1\over I_1}\overline{ M}_{83} - \overline{ R}_{83}
\right).       \label{gai3}
\eeq
The main difference between  \qeq{gai2} and \qeq{gai3} is that the
lowest order term for $g_{\rm A}^{(0)}$ is purely anomalous, whereas the
corresponding term for the $g_{\rm A}^a$ is non-anomalous.
 We come back to this
point, when we make the comparison with the Skyrme model in Sect. VI.

\section{Numerical Results for Axial Currents}

The three different measurements of the spin asymmetry in
the polarized lepton--nucleon deep inelastic scattering
\cite{bolo:EMC}\nocite{bolo:SMC}\nocite{bolo:SMC94}--\cite{bolo:E142}
have been recently
reexamined by Ellis and Karliner \cite{bolo:elka2}.
The message of their work is
that whereas the Bjorken sum rule \cite{bolo:BJ}
is in agreement with the data, the
Ellis--Jaffe sum rule \cite{bolo:ej,bolo:ej2} is violated and the
results read finally:
\beq
g_A^{(0),exp}=0.24\pm 0.09,~~~~g_A^{(8),exp}=0.35\pm 0.04~~{\rm
and}~~g_A^{(3),exp}=1.25.
\label{eq:g083exp}
\eeq

In this section we discuss our numerical results for the three axial
decay constants \gt, \go~ and \gs~ including the strange mass
corrections. They are summarized in  \qtab{tabg3}, \qtab{tabg80} and
\qtab{cont}. Our final values for a constituent quark mass $M=420\MeV$
are given by
\beq
\Gs=0.37,~~~~\Go=0.31~~{\rm and}~~\Gt=1.38.
\label{eq:g083}
\eeq

In \qtab{tabg3}  the difference between SU(2) and SU(3)
results for \gt~ can be seen
in each order of the $1/\Nc$ expansion. Obviously  the
lowest
order contribution $(\Omega^0)$ in SU(3) is significantly smaller than in
SU(2)
due to the fact that the SU(3) expectation value of the corresponding
$D$-function $D_{33}$ is only $70\%$ of the SU(2) value.
The anomalous contribution of Eq.(\ref{g15}) linear in $\Omega$
which is non-zero only in
the SU(3) case acts as a substitute for this group-theoretical reduction.
Indeed, it leads to an almost exact readjustment of
the SU(3) value  to the SU(2) one.
For our preferred value  of $M=423$~MeV from the hyperon spectra
and $\ms=180$~MeV
it pushes the leading order SU(3) result up to
$g_{\rm A}^{(3)}\simeq0.84$ .
These two values of the model parameters
$M$ and $\ms$ will be used in the following
discussion of the numerical results.
Due to the presence of the
quantities $Q_{bc}$ from the explicit time-ordering the SU(2), as well as
the SU(3) results, have  corrections
linear in the rotational velocity.  These  conceptually new terms
have no counterparts in the ordinary
Skyrme model.  Similarly to the old non-relativistic quark model
estimates of the $1/\Nc$ correction, i.e.  $g_{\rm A}^{(3)}=\Nc/3+2/3$, these
new terms turn out to be of the order of 50\% of the leading term. For
SU(2) they push $g_{\rm A}^{(3)}$ from $0.84$ to $1.15$ and in SU(3) they
give the final value of 1.31.  Note that the latter value is obtained
with regularized {\it time-ordered} quantities $Q_{bc}$.
In the present work we have derived $Q_{bc}$  from the
 proper-time regularized real
part of the EEA.  In \qref{ab9}, where the regularization was neglected as
a first approximation
\footnote{Note that  for this purpouse the
functional trace $\Sp$ had
to be changed to $\Spto$ which has different properties in time indices.},
the sea part of the quantities $Q_{bc}$ made a $\simeq~30\%$
contribution to the total value of the $Q_{bc}$.
Although the $Q_{bc}$ are finite it should be regularized since it is a
quantity coming from the real part of the EEA. This is done in the
present paper and as a result
with the inclusion of
a regularization function the sea contribution is less than
$3\%$
(for $M=423\MeV$, see
\qtab{tabMpmM45}).

Various contributions from the strange
quark mass (kinematical, dynamical and wave function) increase the value
of $g_{\rm A}^{(3)}$ of about $\simeq5\%$ up to $g_{\rm A}^{(3)}=1.38$, such
that the experimental value
$g_{\rm A}^{(3),exp}=1.25$ is overestimated only by $\simeq~10\%$.

It has to be stressed that this is in contrast to all calculations
within the purely pseudoscalar \quref{psw} or pseudoscalar and  vector
Skyrme model
\cite{bolo:pw2,bolo:pw1}, in which $g_{\rm A}^{(3)}$ is underestimated by
$\simeq~30\%$.  That this significant  difference is  due to the
presence of the new terms from the time-ordering
of the functional trace is most clearly
evident from \qtab{cont}.  There the flavour contributions to the axial
current are given for the Skyrme model and the NJL model
with and without the time-ordered (T) corrections.
Without the  new corrections the NJL model resembles very much the
numerical results of the SU(3) Skyrme model with vector mesons.
This was already noted at the level of the collective hamiltonian
for the mass splittings in \qref{ab4} and here can be seen numerically for
the axial currents with a high accuracy.

Apart from $g_{\rm A}^{(3)}$ we list also the values for $g_{\rm A}^{(0)}$,
partially given already in \qref{ab6},
and $g_{\rm A}^{(8)}$.
The spin of the proton, that is carried by the quarks, and is equal to the
matrix element of the flavour singlet axial vector current,
has no  contribution in the order $\Omega^0$, but gets
a first non-vanishing contribution in the linear order of $\Omega$.
This, as can be seen from \qtab{tabg80}, is also the dominating
contribution which gets only a very small strange mass correction from
the kinematical  or dynamical parts. The corrections from the
wave-function vanish in the linear order of $\ms$, because the
spin operator is diagonal in space of the higher SU(3)
representations. For $M=423$~MeV, the theoretical value of
$g_{\rm A}^{(0)}\simeq0.37$ is a little bit above the experimental error bars,
nevertheless one has to keep  in mind that
the analysis of the experimental data is still under debate
\quref{elka2}.

Experimental extraction of $g_{\rm A}^{(8)}$  from the hyperon
semi-leptonic decays depends on how the  strange quark mass corrections
to the SU(3) symmetric result are taken into account. Therefore the
experimental error bars on this quantity may be at present too small.
In the present calculation we obtain  $g_{\rm A}^{(8)}=0.31$ to be
compared
with the  'experimental' number of \quref{elka2}  $g_{\rm
A}^{(8),exp}=0.35\pm0.04$.

So from our calculations one can conclude that for the 'fixed' mass of
$M=423$~MeV and
$\ms=180$~MeV
all three axial vector coupling constants are  quite close to the
experimental values of \quref{elka2}. From Fig.~6  it can be seen that
for  larger mass of $M\simeq 550$~MeV, $g_{\rm A}^{(3)}$ and $g_{\rm A}^{(0)}$
almost coincide with the experimental values, whereas
$g_{\rm A}^{(8)}$, having  relative large negative strange quark mass
correction,
deviates from  the central value $0.35$. However, ignoring the
large $\ms$
corrections  (as it is usually assumed in the analysis of the hyperon
semileptonic decays \quref{sssw2}), even the value of $g_{\rm A}^{(8)}$
coincides with the experimental value of \qref{elka2}.

In our view there is no reason to talk of a spin crisis of the nucleon.
In the present approach the proton is treated as a many-body system of
valence and sea quarks and the data of the EMC-experiments are basically
reproduced.

\section{Comparison with the  Skyrme Model}

Now we want to compare our results with the Skyrme model,
which can be  regarded as a large constituent quark mass limit of the
NJL model \quref{megrgo},
or -- equivalently -- the large soliton size limit.
It was already mentioned in \qref{ab3} that the formula for the mass
splitting of the present SU(3) NJL model is much  more subtle than
the corresponding expression for the pseudoscalar Skyrme model.
This is  because certain important anomalous
terms dominated by the valence quarks
are missing in the effective action of the Skyrme model.  This has
been  cured
in the Skyrme model only by introducing vector mesons
\cite{bolo:pw1,bolo:pw2}. Therefore we
will focus here on
the Skyrme model, in which vector mesons and in addition kaon
fluctuations and the gauged Wess-Zumino term are added.
Then the collective operator has  the following structure \quref{pw2}:
\bea  {\hat g}_{\rm A}^3 &=& a_1 D_{33} + a_2 d_{3aa} D_{3a} R_a
            + a_3 D_{38}   + a_4 d_{3aa} D_{3a} D_{8a}  \nn
      & &       +
              a_5  D_{33} (1- D_{88})  +  a_6 D_{38} D_{83}
\label{ga2}
\eea
which corresponds effectively to the expression for the NJL model.
%
Although the origin of the various coefficients are quite different in the NJL
and Skyrme model,
%
both
approaches give effectively the same operator structure for $g_{\rm A}$.

However one should stress here two important differences: first of all the new
corrections linear in $\Omega$
which arise due to the {\it time-ordering} within the fermion loop
vanish in the Skyrme model identically.
The Skyrme model is based upon a local lagrangian density which, apart
 from the Wess-Zumino term, is even in time derivatives and therefore
the {\it spatial} components of the axial currents are also even not
allowing for terms linear terms in $\Omega$. Second, even if one restricts
oneself to the terms not including the corrections due
to the {\it time-odering} (local limit) the contribution of the valence
quarks in the present model makes our results qualitatively very
different from the ones of the Skyrme model.
The coefficient $a_2$ e.g. is in the Skyrme model with purely pseudoscalar
mesons
dominated by the induced kaon fluctuations \quref{psw}, which are
neglected  in the present NJL model.
If the vector mesons are included in the Skyrme approach the situation
does not
change qualitatively \cite{bolo:pw2}.
In the Skyrme model   $a_2$  gives  only a $10\%$ contribution to
$g_{\rm A}$
\cite{bolo:psw}, whereas the dominating valence contribution to $M_{44}$
(compare
Tab. \ref{tabM3M44}) in the NJL-model  gives almost a $30\%$
contribution to
$g_{\rm A}$ if the terms due to the {\it time-ordering} are neglected.
The fact that the total values for $g_{\rm A}$ in the {\it local limit}
of the present
approach and in
the vector meson pseudoscalar Skyrme model
in Ref.\cite{bolo:psw} roughly coincide  hinges also on the rescaling
procedure for the parameter {\it e} in Ref. \cite{bolo:psw}, which tends to
increase
$g_{\rm A}$.
The Wess-Zumino contributions to $g_{\rm A}$ in the Skyrme approach
play a minor role, in the NJL model the anomalous part of the action,
containing the WZ term in lowest order of the gradient expansion, gives
$\simeq 1/4$ of the total amount.

\section{Summary and Discussion}

In this paper we have investigated
mass corrections to the hyperon masses and to
the axial currents  of
the semibosonized SU(3) Nambu-Jona-Lasinio model.
Moreover we have extended our recent analysis of the corrections
to the axial currents which appear due to the {\it time-ordering}
of the quark loop and semiclassical quantization \quref{ab9}
to the case of the regularized effective action.
In the semibosonized NJL model
baryons are understood as solitonic solutions of the classical
equations of motion.  However the solitons do not carry proper quantum
numbers and  the semiclassical  quantization procedure has to be
applied in order to describe the mass
splittings
 within the strange baryon   multiplets.  This treatment is based on
introducing time-dependent rotations in the direction of the zero modes
followed by a canonical quantization  of the collective
coordinates of the rotation matrix. Since these zero modes
contribute significantly to the mass splittings \quref{ab4},
it was a challenging task to look at the axial currents
which can be related to the recent measurements of the spin structure
functions \cite{bolo:elka2,bolo:SMC,bolo:SMC94,bolo:EMC}.

First the model parameters were fixed by looking at the hyperon mass
splittings
up to the terms quadratic in $\ms$. These are reproduced with unexpectedly
high accuracy and point towards a constituent quark
mass of
$M\simeq 420$~MeV. We have also explicitly shown that the wave
function corrections and the corrections
 due to the expansion of the effective action are comparable and therefore
it is somehow inconsistent to perform only the Yabu-Ando diagonalization
of the first order hamiltonian.

Second  we considered the axial vector currents with the inclusion
of the
linear corrections in  the rotational velocity. The new contributions
which appear due to the {\it time-ordering} of the quark loop and
semiclassical quantization have been shown explicitly to come from the
real part of the effective action. If the proper-time regularization
is implemented they are dominated by the contribution of the valence
quarks. In the SU(3) model there are also other contributions linear
in $\Omega$ which come from the imaginary part of the effective action
and as such do not require regularization. They are also dominated by
the valence contribution. This concerns the leading term of \gs,
which vanishes in the pure pseudoscalar Skyrme model, whereas  it
is non-vanishing (however small) in the present model in rough agreement
with experiment.

The expression for $g_{\rm A}^{(3)}$  has a
$\simeq25\%$ rotational contribution from the imaginary part of the
effective
Euclidean action, which vanishes  in the SU(2) case and which can
be related to Witten's formula for the axial vector
current
 from the Wess-Zumino effective action.  Moreover
it has a
$\simeq30\%$  contribution due to
the  explicit
{\it time-ordering}
$(Q_{bc})$ of the collective operators. These terms are not present in local
theories like the Skyrme model. In the
present model they are entirely due to the {\it non-locality} of the
fermion
determinant. Performing the gradient expansion of these quantities, it can
be shown that these terms have the same mesonic structure  as
the lowest order term, such that one can define a profile
independent renormalization of  the lowest order $g_{\rm A}^a$.
This is similar to recent findings of Dashen and Manohar
\quref{dama2} within large-$\Nc$  QCD  and to the old non-relativistic
quark model result of
$g_{\rm A}^{(3)}=(\Nc+2)/3$.

Quantitatively
despite the fact that the  lowest order SU(3) result
is reduced by a group theoretical factor of $0.7$ with respect to the
SU(2) case the new
time-ordering and
anomalous contributions push the total value of $g_{\rm A}^{(3)}$ upwards.

Next we have considered corrections linear in the
strange quark mass. They are consistently derived from
the expansion of the effective action, the quantization condition as
well as from the higher
representations of the wave function in the spirit of the Yabu-Ando
diagonalization.
However the effect on $g_{\rm A}^{(3)}$ is not large and finally one
ends up with
 $g_{\rm A}^{(3)}=1.38$ for $M=423$~MeV, which is only  $\simeq~10\%$
above the experimental value of $g_A^{(3),exp}=1.25$. Here it should be
stressed,
that such nice agreement  was never obtained within the pseudoscalar or
pseudoscalar vector Skyrme model \cite{bolo:psw,bolo:pw2}.
This qualitative and quantitative difference comes from the new
$1/\Nc$ corrections of the time-ordering from the non-local structure of
the present NJL model.

The
$g_{\rm A}^{(0)}$ exists already in SU(2) and the only effect in
SU(3) is a small shift due to the  finite symmetry breaking  $\ms$.
This is in contrast   to $g_{\rm A}^{(8)}$, which vanishes in the SU(2)
case, and
which  in SU(3) gets entire contribution  from the rotation and from
the strange quark mass.
In the chiral limit the  values for $g_{\rm A}^{(8)}$ and $g_{\rm A}^{(0)}$ are
quite
close to each other, as suggested in \quref{bekame}, however the strange
mass corrections reduce the value of  $g_{\rm A}^{(8)}$ by $\simeq~25\%$,
whereas
the explicit symmetry breaking  has almost no influence on  $g_{\rm A}^{(0)}$.
This holds at least  if we take all linear $\ms$ corrections into
account  and even  the $\ms^2$-corrections, which can be calculated
in this framework  from the non-symmetric wave-functions
\quref{ab5}.
The final values $g_{\rm A}^{(0)}\simeq 0.37$  and
$g_{\rm A}^{(8)}\simeq 0.31$  for
$M=423$~MeV and $\ms=180$~MeV
are to be compared with the experimental data extracted from the recent
Ellis and Karliner analysis \quref{elka2}, i.e.
$g_{\rm A}^{(0),exp}\simeq 0.24\pm0.09$ and
$g_{\rm A}^{(8),exp}\simeq 0.35\pm0.04$. Apparently the theoretical values
are only slightly outside.

Qualitatively the following can be said:
 $g_{\rm A}^{(0)}$, which represents
the part of the proton spin carried by the quarks, gets a
non-vanishing expectation value  entirely due to the anomalous part
of the EEA.
In a constitituent quark model, when the total spin of the proton is
entirely carried by three quarks, this $g_A^{(0)}$ equals {\it one}.
The present model gives values close to experiment since the proton is
treated entirely as many body system rotating in spin-isospin space.
Hence one cannot attribute the spin of the proton to single elementary
particles but only to the whole system.
Hence the fact that in agreement with the experiments only a
fraction of
about $20-30\%$ of the nucleon spin is carried by the quark-spins is not
surprising at all in the present model.
One should note that
in the Skyrme models a non-vanishing value of $g_A^{(0)}$
can be obtained only by adding vector mesons for the anomalous part
\cite{bolo:jjmps,bolo:sssw2,bolo:pw2}.


Altogether the picture which emerges is quite satisfactory. Mass
splittings are accurately reproduced and axial currents are in good
agreement with the experimental data if rotational $1/N_c$ corrections
are taken into account. In particular the spin of the proton
originates in this model to about $35\%$ from the spin of the quarks, a
number being in reasonable agreement to the world data reported by SMC
\quref{SMC94}.
Together with the numerical results for the Gottfried sum \quref{ab13}
the model provides a good reproduction of experimenta data so far.

On purely theoretical side the presence of
the new terms linear in $\Omega$
calculated in this paper poses a serious problem to  effective
meson theories like the Skyrme model, where such terms vanish identically.
Another theoretical question which deserves a comment is the convergence of
the expansion in $\Omega$. The large size of the corrections
calculated in this paper, although expected from the quark model
calculations, poses a problem of reliability of the numerical
results. One way to tackle this problem would be to calculate
the $\Omega^2$ corrections to the axial currents. This is the
highest power of $\Omega$ one can think of, since the collective
hamiltonian itself is of that order. Despite the technical
difficulties in calculating these terms the preliminary estimates
indicate that they are not negligible\footnote{M. Wkamatsu,
private communication}. Therefore one has to conclude that the
expansion in $\Omega$ is slowly convergent. Moreover the
formalism of the collective quantization has to be revised if
one wants to include terms higher than $\Omega^2$.
In addition despite the fact that mass splittings are well reproduced
the absolute energies provide  still some problems which are associated
with zero-point corrections \cite{bolo:pob,bolo:ab4} and boson
fluctuations.
These
questions are certainly beyond the scope of this paper, where we
had to content ourselves with the linear corrections alone.

\vspace{1 cm}

\acknowledgements

The work was partially supported  by {\it Alexander von
Humboldt Stiftung} and {\it Polish Research Grant} KBN~2.0091.91.01 (M.P.),
{\it Graduiertenstipendium des Landes NRW} (A.B.),
and the
{\it Bundesministerium fuer Forschung und Technologie} (BMFT, Bonn).
One of us  (A.B.) would also like to thank the
{\it Ruth und Gerd
Massenberg-Stiftung}   and  {\it The Institute of Physics} (IOP)
for financial  support.
Discussions with  Ch. Christov (Bochum),
D. Diakonov, V. Petrov, M. Polyakov
(Petersburg),
M. Wakamatsu (Osaka) and T. Watabe (Tokyo)  are acknowledged.
\lb
The material of this work was presented at the {\it International
Workshop on the Quark Structure of Baryons} in Trento, Italy, at the
new {\it European Centre for Theoretical Studies} (${\rm ECT}^*$),
October 4-15, 1993.
\vfill\eject

\begin{table}
\caption{Moments of inertia for different constituent masses}
\label{tabIKN}
\begin{center}
\begin{tabular}{ccccccccr}
$M$&$\Sigma$~[SU(2)]&$I_1$ & $I_2$ & $K_1$ & $K_2$ &$N_0$&$N_1$&$N_2$
\\
{}~$$[MeV] & [MeV] & [fm] & [fm] & [fm] & [fm] & [fm] & [fm] & [fm]   \\
\hline
 363. & 60.32& 1.512& 0.720& 0.606& 0.372& 0.765& 0.647& 0.496  \\
 395. & 58.14& 1.285& 0.618& 0.438& 0.290& 0.704& 0.500& 0.408  \\
 419. & 56.14& 1.178& 0.569& 0.369& 0.255& 0.668& 0.438& 0.370  \\
 423. & 55.52& 1.156& 0.560& 0.357& 0.250& 0.658& 0.426& 0.362  \\
 465. & 51.86& 1.015& 0.496& 0.276& 0.210& 0.599& 0.349& 0.311  \\
\end{tabular}
\end{center}
\end{table}
\begin{table}
\caption{Coefficients of Eqs.(22)  as functions of
constituent mass $M$ for $\ms=180$~MeV }
\label{tababcdfgh}
\begin{center}
\begin{tabular}{cccccccc}
  M   &   A     &   F     &   D     &  G     &   B     &   C     & H  \\
\hline
   363. & 489.04 & 427.02 &  94.67 &   1.96 & 450.89 & 510.24 &   6.70 \\
   395. & 481.32 & 399.60 &  97.41 &   1.70 & 449.48 & 457.80 &   5.56 \\
   419. & 469.60 & 381.66 &  97.17 &   1.51 & 441.70 & 427.02 &   4.86 \\
   423. & 465.57 & 377.47 &  96.79 &   1.48 & 438.60 & 420.32 &   4.70 \\
   465. & 441.39 & 349.75 &  94.91 &   1.25 & 419.82 & 376.11 &   3.72 \\
\end{tabular}
\end{center}
\end{table}
\begin{table}
\caption{Different contributions to coefficients of Eqs.(22)
for a constituent mass $M=423$ MeV and for $\ms=180$~MeV }
\label{tabABCDFGH}
\begin{center}
\begin{tabular}{cccccccc}
  & $O(\ms)$ &\multicolumn{4}{c}{$O(\ms^2)$} & total & exp. \\
  \cline{3-6}
  &        &    kin.  &   dyn.    &   w.f. & total  &         &    \\
  \hline
 A &  546.10 &   10.94 &  -64.64 &   -0.15 &  -53.85 &  492.25 &  -- \\
 B &  546.10 &   10.85 &  -64.55 &  -53.81 & -107.50 &  438.60 &  --  \\
 F &  381.20 &    1.18 &  -27.67 &   22.76 &   -3.73 &  377.47 &  379.00 \\
 D &  120.76 &   -0.02 &  -11.78 &   -0.07 &  -11.87 &  108.89 &   86.00 \\
 C &  348.16 &    1.20 &  -19.97 &   90.92 &   72.15 &  420.32 &  422.00 \\
 G &    0.00 &    0.61 &   -0.66 &    1.53 &    1.48 &    1.48 &    6.75 \\
 H &    0.00 &   -0.29 &    0.41 &    4.57 &    4.70 &    4.70 &    2.20 \\
\end{tabular}
\end{center}
\end{table}
\begin{table}
\caption{
Quantities  $M_3$ and $M_{44}$
for the SU(3) Nambu--Jona-Lasinio model in dependence on
the constituent quark mass $M$
}
\label{tabM3M44}
\begin{center}
\begin{tabular}{cccccccc}
  && $M_3$ &&&&  $M_{44}$~[fm] & \\
\cline{2-4}  \cline{6-8}
$M$~[MeV] & val & sea  & tot& &val   & sea    & tot  \\
\hline
 363     & -2.293 &-0.468 & -2.761 & & -0.414 & -0.011 & -0.425  \\
 395     & -2.235 &-0.384 & -2.619 & & -0.326 & -0.012 & -0.338  \\
 419     & -2.209 &-0.316 & -2.525 & & -0.288 & -0.012 & -0.301   \\
 423     & -2.205 &-0.307 & -2.511 & & -0.283 & -0.013 & -0.295   \\
 465    & -2.173 &-0.203 & -2.377 &  &-0.238 & -0.014 & -0.251    \\
\end{tabular}
\end{center}
\end{table}
\begin{table}
\caption{
Quantiyies  $Q_{12}=-2iQ_{-+}$ and  $Q_{45}=i{\tilde Q}_{45}$
for the SU(3) Nambu--Jona-Lasinio model in dependence on
the constituent quark mass $M$
}
\label{tabMpmM45}
\begin{center}
\begin{tabular}{cccccccr}
  && $Q_{-+}$~[fm] &&&&  ${\tilde Q}_{45}$~[fm] & \\
\cline{2-4}  \cline{6-8}
$M$~[MeV] & val & sea  & tot&& val   & sea    & tot  \\
\hline
 363       & .408   &.025   &.433  & & -.410   & -.023   & -.433    \\
  395      & .317   &.021   &.339  & &-.318   & -.020   & -.338    \\
     419   & .279   &.019   &.298  & &-.279   & -.018   & -.297    \\
     423   & .272   &.019   &.291  & &-.273   & -.017   & -.290    \\
     465   & .226   &.016   &.242  & &-.226   & -.015   & -.241    \\
\end{tabular}
\end{center}
\end{table}
\begin{table}
\caption{
Quantities  $\overline{ M}_{83}$ and  $\overline{
R}_{83}$
for the SU(3) Nambu--Jona-Lasinio model in dependence on
the constituent quark mass $M$
}
\label{tabMR83}
\begin{center}
\begin{tabular}{cccccccr}
  && $\overline{ M}_{83}$~[fm] &&&&  $\overline{ R}_{83}$~[fm] & \\
\cline{2-4}  \cline{6-8}
$M$~[MeV] & val & sea  & tot&& val   & sea    & tot  \\
\hline
 363  &-0.683 &-0.015 &-0.699 &&-0.293  &-0.094 & -0.387  \\
  395     &-0.500 &-0.016 &-0.516 &&-0.150  &-0.090 & -0.240  \\
 419  &-0.422 &-0.016 &-0.438 &&-0.095  &-0.091 & -0.186  \\
 423  &-0.409 &-0.016 &-0.425 &&-0.086  &-0.091 & -0.177  \\
 465  &-0.316 &-0.017 &-0.333 &&-0.025  &-0.090 & -0.115  \\
\end{tabular}
\end{center}
\end{table}

\begin{table}
\caption{
Quantities   $R_{44}$  and  $\overline{ R}_{38}$
for the SU(3) Nambu--Jona-Lasinio model in dependence on
the constituent quark mass $M$
}
\label{tabR44R38}
\begin{center}
\begin{tabular}{cccccccr}
  && $R_{44}$~[fm] &&&&  $ \overline{ R}_{38}$~[fm] & \\
\cline{2-4}  \cline{6-8}
$M$~[MeV] & val & sea  & tot&& val   & sea    & tot  \\
\hline
 363  &-0.253 & -0.023 & -0.277  & & 0.074 & -0.049 &  0.024   \\
  395     &-0.178 & -0.029 & -0.207  & & 0.083 & -0.066 &  0.017   \\
 419  &-0.148 & -0.030 & -0.179  & & 0.086 & -0.073 &  0.012   \\
 423  &-0.144 & -0.031 & -0.175  & & 0.087 & -0.075 &  0.012    \\
 465  &-0.111 & -0.033 & -0.143  & & 0.090 & -0.083 &  0.006   \\
\end{tabular}
\end{center}
\end{table}
\begin{table}
\caption{
The  axial vector coupling constant
$g_{\rm A}^{(3)}$ for the SU(3)   Nambu--Jona-Lasinio model in
dependence on
the constituent quark mass $M$. The strange current quark mass is chosen
as $\ms=180$~MeV. The final model predictions are given by
$g_A^{(3)}(\Omega^1)$ in SU(2) and $g_A^{(3)}(\Omega^1,m_s)$ in SU(3).
The experimental value is given by
$g_A^{(3),exp}=1.25$.
}
\label{tabg3}
\begin{center}
\begin{tabular}{ccccccc}
  &\multicolumn{2}{c}{SU(2)}& &\multicolumn{3}{c}{SU(3)} \\
\cline{2-3} \cline{5-7}
$M$~[MeV]   &   $g_{\rm A}^{(3)}(\Omega^0)$   &   $g_{\rm
A}^{(3)}(\Omega^1)$  & &  $g_{\rm A}^{(3)}(\Omega^0,\ms^0)$
 &  $g_{\rm A}^{(3)}(\Omega^1,\ms^0)$  &  $g_{\rm
     A}^{(3)}(\Omega^1,\ms^1)$ \\
\hline
     363 &   0.920 & 1.302&  & 0.644  & 1.482  & 1.603 \\
  395    &   0.873 & 1.224&  & 0.611  & 1.381  & 1.473 \\
     419 &   0.841 & 1.179&  & 0.589  & 1.328  & 1.407 \\
     423 &   0.837 & 1.173&  & 0.585  & 1.314  & 1.380 \\
     465 &   0.792 & 1.109&  & 0.554  & 1.250  & 1.308 \\
\end{tabular}
\end{center}
\end{table}
\begin{table}
\caption{
The  axial vector coupling constant
$g_{\rm A}^{(0)}$ and $g_{\rm A}^{(8)}$ for the SU(3)
Nambu--Jona-Lasinio model in dependence of
the constituent quark mass $M$. The strange current quark mass is chosen
as $\ms=180$~MeV.
The final model predictions are given by
$g_A^{(0)}(\Omega^1,m_s)$  and $g_A^{(8)}(\Omega^1,m_s)$.
The experimental values are given by
$g_A^{(8),exp}=0.35\pm0.04$ and
$g_A^{(0),exp}=0.24\pm0.09$ (Ellis and Karliner [50]).
}
\label{tabg80}
\begin{center}
\begin{tabular}{ccccccc}
$M$~[MeV]   &   $g_{\rm A}^{(8)}(\Omega^0,\ms^0)$
  &   $g_{\rm A}^{(8)}(\Omega^1,\ms^0)$
  &  $g_{\rm A}^{(8)}(\Omega^1,\ms^1)$  &  &   $g_{\rm
A}^{(0)}(\Omega^1,\ms^0)$
  &  $g_{\rm A}^{(0)}(\Omega^1,\ms^1)$     \\
\hline
    363    &   0.159 & 0.443   &0.328   &  &  0.462 & 0.475    \\
   395     &   0.151 & 0.408   &0.316   &  & 0.401 & 0.409    \\
      419  &   0.145 & 0.389   &0.309   &  & 0.371 & 0.377    \\
      423  &   0.144 & 0.385   &0.308   &  & 0.364 & 0.370    \\
      465  &   0.137 & 0.363   &0.301   &  & 0.328 & 0.331    \\
\end{tabular}
\end{center}
\end{table}
\begin{table}
\caption{Various contributions to the axial vector current of the
proton in terms of u,d and s quarks. A comparison is made between
the Skyrme model with vector mesons [24] (Skyrme,vector),
the NJL model
without (NJL,scalar) and with the
{\it time-ordered} corrections
of this paper (NJL,scalar,T). In the last column experimental values
 from Ellis and Karliner [50] are given.}
\label{cont}
\begin{center}
\begin{tabular}{ccccc}
   &  Skyrme (vector)
  &   NJL (scalar)
  &   NJL (scalar,T)
  &   'experiment'  \\
\hline
$\Delta u$  & ~0.63   &  ~0.64     & ~0.902   & ~0.806  \\
$\Delta d$  & -0.31   &  -0.24     & -0.478   & -0.444  \\
$\Delta s$  & -0.03   &  -0.02     & -0.054   & -0.120  \\
\end{tabular}
\end{center}
\end{table}
\begin{table}
\caption{
 Matrix elemnets    of the operators for $g_{\rm A}^a$ in the proton state
 with spin up,
 where the index $i$ is always running from 1 to 3
and $b$ from 4 to 7.
}
\label{tabDDJ}
\begin{center}
\begin{tabular}{ccccccccc}
         $D_{33}$   & $D_{38}$   &  $D_{88}$   &
        $d_{3bb} D_{3b} \hat{J}_b $   &  $d_{3bb} D_{8b} \hat{J}_b $   &
        $d_{3bb} D_{3b} D_{8b}$ & $d_{3bb} D_{8b} D_{8b}$

\\
\hline
\smallskip
 $-7/30$   & $ \sqrt{3}/30$ & $3/10$  &$ 7/60$ &$ 1/(20\sqrt{3})$ &
$-11/(90\sqrt{3})$
& 1/30 \\
\hline\hline
\smallskip
  $D_{38} D_{83}$  &
 $D_{33} D_{88}$ &  $D_{83}$   &   $D_{83}D_{88}$   & $D_{3i}R_i$ &
$D_{3i}D_{8i}$ &  $D_{3b}D_{8b}$
\\
\hline
 $-1/45 $ &
 $-4/45$  &      $-\sqrt{3}/30$ &   0   & $7/20$  & $\sqrt{3}/45$
&$-\sqrt{3}/45$
\\
\end{tabular}
\end{center}
\end{table}
\appendix

\section{Semiclassical Quantization}

In this Appendix
\footnote{We are
greatful to P. Pobylitsa for discussion and for making clear to us many
subtleties of the group theory}
we collect all essential ingridients of the
semiclassical quantization scheme.
For the purpose of simplicity we will consider only $L_0$ without
mass corrections. The generalization to include $\ms$ terms is
straightforward. To this end one defines curvlinear angular velocities:
\beq
A^{\dagger}\dot{A}=\frac{i}{2} \Omega_i \lambda_i
  =\frac{i}{2} \dot{\theta}_k e_{ki}(\theta) \lambda_i,
\eeq
where the {\it vielbeins} fulfil:
$e_{ki}\;\overline{e}_{kj}=\delta_{ij}$ and
$e_{ik}\;\overline{e}_{jk}=\delta_{ij}$
(the {\it bar} denotes the inverse {\it vielbein}).
The generalized momenta are defined as:
\beq
\pi_i=\frac{\partial L_0}{\partial \dot{\theta}_i}=
\dot{\theta}_j e_{jm} I_{mn} e_{in}-c\;  e_{i8}, \label{eq:pii}
\eeq
where
$c=\Nc/\sqrt{12}$.
Here one postulates canonical commutation rules:
\beq
[{\hat {\pi}}_i,{\hat {\theta}}_j]=-i \delta_{ij}
\eeq
and then the differencial representation for the operator
${\hat {\pi}}$ is, as
usual: $
{\hat {\pi}}_i=-i \partial / \partial \theta_i \equiv -i \partial_i$.

It is convenient to define the quantities $J_i$:
\beq
J_j=\overline{e}_{ij} \pi_i= \Omega_{m} I_{mj} - c\; \delta_{8j}.
\eeq
The normal quantization procedure is here slightly subtle, since the
tensor
$I$ is not invertible and hence $J_8=-c$ is a constraint rather than the
dynamical variable. To circumvent this difficulty let us introduce a small
moment of inertia $I_{88}=\varepsilon$ and then take $\varepsilon
\rightarrow 0$. Then one can proceed in a normal way, inverting
Eq.(\ref{eq:pii}) for velocities and calculating the hamiltonian:
\beq
H^{(0)}=M_{\rm cl} + \frac{1}{2I_1}\sum\limits_{i=1}^3 J_i^2 +
  \frac{1}{2I_2}\sum\limits_{i=4}^7 J_i^2 +
\frac{1}{2 \varepsilon} (J_8+c)^2.
\eeq
In the limit $\varepsilon \rightarrow 0$ one recovers the constraint.

Upon quantization $J_i$ are promoted to the  operators:
${\hat J}_j=-i\; \overline{e}_{ij} \partial_i$.
It is straigthforward, but tedious to calculate the commutation relations
for ${\hat J}_i$'s. To this end one has to use the following identities:
\bea
\partial_j e_{ik} -\partial_i e_{jk}&  = & e_{im} e_{jn}\; f_{mnk},
\nonumber \\
\overline{e}_{jm} \partial_j \overline{e}_{in} -
\overline{e}_{jn} \partial_j \overline{e}_{im}
& = &  - \overline{e}_{ik} f_{kmn}.
\eea
Finally one arrives at:
\beq
[{\hat J}_i,{\hat J}_j]= i f_{ijk} {\hat J}_k.
\eeq
The next question is how to interpret operators ${\hat J}_i$. There are two
global symmetries of the rotation matrix $A$ which do not change the
Lagrangian $L_0$:
\bea
{\rm left:~~~} A\rightarrow e^{i\frac{1}{2}\vec{\varphi}{\vec{\lambda}}}\;A
&{\rm ~~~~and~~~}&
{\rm right:~~~} A\rightarrow A\;e^{-i\frac{1}{2}\vec{\varphi}
\vec{{\lambda}}}.
\eea
In the case of the right multiplication only $\lambda_{1,2,3}$ enter, since
due to the form of the {\it hedgehog} Ansatz the right multiplication has to
commute with the right multiplication by $\lambda_8$. Then it becomes
clear that the right symmetry corresponds to rotations, since it can be
undone by the rotation of vector $\vec{n}$ entering the {\it hedgehog}
Ansatz. On the contrary the left multiplication is just a global
symmetry of the Lagrangian and it can be interpreted as a flavor symmetry.
To calculate the generators of these symmetries one uses the Noether
construction  promoting $\varphi$ to a time dependent quantity. Then,
for infinitesimal transformations, one gets that:
\[{\rm left:~~~} \Omega_i \rightarrow \Omega_i+\dot{\varphi}_j D_{ji}
{\rm~~~and~~~}
{\rm right:~~~} \Omega_i \rightarrow \Omega_i-\dot{\varphi}_i \]
and subsequently:
\bea
{\rm left:~~~}
-\frac{\partial}{\partial\dot{\varphi}_i} L_0=-D_{ij} J_j\equiv T_i
&{\rm ~~~and~~~}&
{\rm right:~~~}
-\frac{\partial}{\partial\dot{\varphi}_i} L_0= J_i.\label{eq:gens}
\eea
Here $D_{ij}\equiv D^{(8)}_{ij}=
1/2\; {\rm Tr}(A^{\dagger}\lambda_i A \lambda_j)$ are the Wigner
matrices in the adjoint representation of the SU(3) group.
We have already shown that upon the quantization $J_i$ are promoted to
the SU(3) generators, however only ${\hat J}_{1,2,3}$
correspond to the symmetry
generators, namely to spin. To evaluate the commutation relations for
${\hat T}_i$ one has to convince onself that:
\[ [{\hat J}_i, D_{aj}]= i f_{ijk} D_{ak}. \]
Then it follows that:
\beq
[{\hat T}_i,{\hat T}_j]= i f_{ijk}{\hat T}_k.
\eeq
Making use of the first of Eqs.(\ref{eq:gens}) and of the ortogonality
relation for the $D$ functions: $D_{ac}D_{bc}=D_{ca}D_{cb}=\delta_{ab}$
one can rewrite $H^{(0)}$ as in \qeq{ham0}.

Baryon wave functions should fulfil the following identities:
\bea
\left[ \exp(i\omega_a \hat T_a)\; \psi \right](A)
= \psi \left(e^{-i\omega_a {\lambda_a}/{2}} A \right),
& &
\left[ \exp(i\omega_a \hat J_a) \; \psi \right](A)
= \psi \left (Ae^{i\omega_a \lambda_a/{2}} \right).
\label{eq:psi1}
\eea
The phase convention in Eq.(\ref{eq:psi1}) is chosen in such a way that
the $\psi$'s are faithful, {\it i.e.}:
\bea
\left[ \exp(i\omega^{(2)}_a \hat T_a)\;\exp(i\omega^{(1)}_a \hat T_a)\; \psi
\right](A)
&=& \psi\left( \left\{ e^{i\omega^{(2)}_a \lambda_a /2} \;
       e^{i\omega^{(1)}_a \lambda_a /2} \right\}^{-1} A \right),
       \nonumber \\
\left[ \exp(i\omega^{(2)}_a \hat J_a)\;\exp(i\omega^{(1)}_a \hat J_a)\; \psi
\right](A)
&=& \psi\left( A\; e^{i\omega^{(2)}_a \lambda_a /2}\;
      e^{i\omega^{(1)}_a \lambda_a /2}  \right). \nonumber
\eea

The problem of constructing the baryon wave functions reduces to the
construction of the functions of  $A$ which
 transform under left and
right rotations as the tensors of the irreducible representations
${\cal T}$ and ${\cal J}$ respectively. In the matrix representation
that means:
\bea
\left[ \exp(-i\omega_a \hat T_a)\;
\psi^{({\cal T J})} \right]_{tj} (A)
&=&
\psi^{({\cal T J})}_{t^{\prime} j} \left( A \right)\;
 D^{({\cal T})}_{t^{\prime} t} \left( e^{-i\omega_a {\lambda_a}} \right),
 \nonumber \\
\left[~ \exp(i\omega_a \hat J_a)~ \;
\psi^{({\cal T J})} \right]_{tj}(A)
&=&
\psi^{({\cal T J})}_{t j^{\prime}} \left( A \right)\;
 D^{({\cal J})}_{j^{\prime}j} \left(~ e^{i\omega_a {\lambda_a}}~ \right).
\label{eq:psi2}
\eea
Replacing LHS in Eqs.(\ref{eq:psi2}) by RHS of Eqs.(\ref{eq:psi1})
and making subbstitution: $A\rightarrow 1$ and
$\exp(i\omega_a \lambda_a)\rightarrow A$ one gets:
\bea
\psi^{({\cal T J})}_{tj} (A) =
c_{t^{\prime}j} D^{({\cal T})}_{t^{\prime} t} ( A^{\dagger} ), & &
\psi^{({\cal T J})}_{tj} (A) =
c_{t j^{\prime}} D^{({\cal J})}_{j^{\prime} j} ( A ) \label{eq:psi3}
\eea
where $c_{t j}=\psi^{({\cal T J})}_{tj}(1)$.

 In order to calculate the constants $c_{t j}$ let us observe that:
\beq
\psi^{({\cal T J})}_{tj} (L^{\dagger}AR) =
\psi^{({\cal T J})}_{t^{\prime} j^{\prime}} \left( A \right)\;
 D^{({\cal T})}_{t^{\prime}t}(L) \;
 D^{({\cal J})}_{j^{\prime}j}(R).\label{eq:LAR}
\eeq

We will now make use of the following identity conecting elements of
the Wigner matrices in representation ${\cal T}$ and its complex
conjugate $\overline{\cal T}$:
\beq
D^{({\cal T})}_{ij}(L) = (-)^{Q(i)-Q(j)}\;
D^{(\overline{\cal T})}_{{-j}{-i}}(L^{\dagger}),
\eeq
where $Q$ denotes the charge of the state: $Y,I,I_3$ and {\it minus}
before
the index $i$ or $j$ denotes the complex conjugate state: $-Y,I,-I_3$.
Putting $A=1$ in Eq.(\ref{eq:LAR}) we we find that
$\psi(L^{\dagger}R)$ is proportional to the product of
$D^{(\overline{\cal T})}(L^{\dagger})\; D^{({\cal J})}(R)$.
On the other hand, following
Eq.(\ref{eq:psi3}), $\psi(L^{\dagger}R)$
  should be proportional
either to $D^{(\overline{\cal T})}(L^{\dagger}R)$ or
$D^{({\cal J})}(L^{\dagger}R)$. That means immediately that
\[ \overline{\cal T} = {\cal J}. \]
One can also easily convince oneself that:
\[ c_{tj} = (-)^{-Q(t)}\delta_{-tj}.   \]
So we get two equivalent expressions for the baryon wave function:
\beq
\psi^{({\cal T})}_{tj} (A) =
(-)^{Q(j)}\; D^{({\cal T})}_{-j t} ( A^{\dagger} )
  =  (-)^{Q(t)}\; D^{(\overline{\cal T})}_{-t j} (A). \label{eq:psi4}
\eeq
Throughout this derivation we have assumed that charges are integer.
Finally making use of the fact that
$D^{({\cal T})}_{ij}(A^{\dagger})={D^{({\cal T})}_{ji}(A)}^{\bf *}$
we get our final expression for the baryon wave function:
\beq
\psi^{({\cal T})}_{(Y~ T~ T_3)\;(Y^{\prime} J~J_3)} (A)
= \sqrt{{\rm dim}\;{\cal T}}\;
(-)^{Y^{\prime}/2+J_3}\;
{\left[<Y,T,T_3 \mid D^{({\cal T})}(A)
\mid -Y^{\prime},J,-J_3>\right]}^{\bf *} ,
\eeq
where the normalization factor has been included. Remember that the
right hypercharge is constrained: $Y^{\prime}=-\Nc/3$

The action of the colletive operators on these wave functions is
straightforward: flavor operators ${\hat T}_a$ and spin oerators
${\hat J}_b$ (also for $b=4 \ldots 8$) act on $\psi$ as the SU(3)
generators
in representation ${\cal T}$ act on the state $(Y,T,T_3)$ and
$(Y^{\prime},J,J_3)$ respectively. The action of the  $D$
functions entering
the collective operators can be calculated with the help of the SU(3)
Clebsch-Gordan coefficients:
\begin{eqnarray}
{\rm dim}\; {\cal T}_{3} \,
\int dA\: { D_{t_3 j_3}^{({\cal T}_3)} (A) }^{\bf *}\,
D_{t_2 j_2}^{({\cal T}_2)}(A)\, D_{t_1 j_1}^{({\cal T}_1)}(A)
&=& \sum\limits_{\gamma}
\left(
\begin{array}{ccl}
{\cal T}_{1} &  {\cal T}_{2} & {\cal T}_{3}^{\gamma} \\
t_{1}    &  t_{2}    & t_{3}
\end{array}
\right)
\left(
\begin{array}{ccl}
{\cal T}_{1} &  {\cal T}_{2} & {\cal T}_{3}^{\gamma} \\
j_{1}    &  j_{2}    & j_{3}
\end{array}
\right),
\label{eq:Dort}
\end{eqnarray}
where $\gamma$ is the degeneracy index.
In \qtab{tabDDJ} we list some of the proton
spin up matrix elements of the
collective operators which enter the expressions for $g_{\rm A}^a$.

\section{Derivation of the Regularization Functions for the Axial
Current}

Here we want to give an explicit derivation of the $\Omega^0$ and
$\Omega^1$ contributions to the axial current. We emphasize the
method of regularization of non-anomalous quantities from the explicit
{\it time-ordering} of the collective  operators within
the proper-time regularization
scheme \quref{schwinger}. Then the real part can be written as:
\beq   {\rm Re}\;  A_i^a(x)  = -\half{\delta\over \delta s_i^a(x)}  \Spto
\int {d             u\over u}\phi   (u) \exp{(-u D^\dagger D)}  \label{aa1}
	  \eeq where \beq D=\partial_t + H + i\Omega_{\rm E} - i s_i^a
\gamma_4\gamma_i\gamma_5 A^\dagger \lambda^a A  \eeq and \beq
D^\dagger=-\partial_t + H - i\Omega_{\rm E} - i s_i^a
\gamma_4\gamma_i\gamma_5
      A^\dagger \lambda^a A,   \label{aa2}  \eeq
such that
\bea
 D^\dagger D &=& -\partial^2+H^2+\Omega_{\rm E}^2
-i[\Omega_{\rm E},H]-i \{\Omega_{\rm E},\partial_t\}
 -is_i^a \{ \gamma_4\gamma_i\gamma_5 A^\dagger\lambda_aA,H \}  \nn
 & & +s_i^a \gamma_4\gamma_i\gamma_5 [A^\dagger\lambda_aA,\Omega_{\rm E}]
  -i  \gamma_4\gamma_i\gamma_5
  [s_i^a A^\dagger\lambda_aA,\partial_t].   \label{aa3}
\eea
Then one has to expand $D^\dagger D$ around  $D_0^\dagger D_0$ with
$D_0=\partial_t + H$, i.e.  one expands in terms of
$s_i^a$ and $\Omega_{\rm E}$. This is done by using
the Schwinger Dyson formula:
\bea  && \exp{(-uD^\dagger D)} = \exp{(-uD_0^\dagger D_0)} \nn
    &&  -u\; \int\limits_0^1 d \alpha
      \exp{(-u\alpha D_0^\dagger D_0)}
     [ D^\dagger D - D_0^\dagger D_0 ]
       \exp{(-u(1-\alpha) D_0^\dagger D_0)}    \nn
  & & +u^2\; \int\limits_0^1 d \beta   \int\limits_0^{1-\beta}  d \alpha
     e^{ -u\alpha D_0^\dagger D_0 }
    [  D^\dagger D - D_0^\dagger D_0 ]
       e^{ -u\beta  D_0^\dagger D_0 }
[  D^\dagger D - D_0^\dagger D_0 ]
      e^{-u(1-\alpha-\beta)  D_0^\dagger D_0} \nn
 & & + \dots.
   \label{aa4}  \eea

In the lowest order $\Omega^0$, one obtains
\bea   {\rm Re}\;  A_i^a(x) &=&-\half{\delta\over \delta s_i^a(x)}  \Spto
	   \int {d  u\over u}\phi   (u) \int d\alpha  \exp{(-u \alpha
	  D_0^\dagger D_0)}  \nn & & \left( u is_i^a \{
    \gamma_4\gamma_i\gamma_5 A^\dagger\lambda_aA,H \} \right) \exp{(-u
	  (1-\alpha) D_0^\dagger D_0)}, \label{aa5}  \eea which after
some simple manipulations gives \qeq{g5e} (see \qref{mego}).

Now we want to consider the $\Omega_{\rm E}^1$ corrections to the current.
Let us
define  $V_1= -i[\Omega_{\rm E},H]-i \{\Omega_{\rm E},\partial_t\}$,
$V_2=-is_i^a\{\gamma_4\gamma_i\gamma_5 A^\dagger\lambda_aA,H\}$,
$V_3=s_i^a \gamma_4\gamma_i\gamma_5 [A^\dagger\lambda_aA,\Omega_{\rm E}]$ and
$V_4=-i\gamma_4\gamma_i\gamma_5[s_i^aA^\dagger\lambda_aA,\partial_t]$.
Consistently in the order $\Omega_{\rm E}^1$ one has to consider combinations
of
$V_1$ and
$V_2$ as well as the single sum  $V_3+V_4$.
Note that it is important to retain $s_i^a$ as time-dependent in
\qeq{aa3}, because
otherwise the two terms $V_3+V_4$  cancel. This can be seen using
$[A^\dagger\lambda_aA,\partial_t]=i[\Omega_{\rm E},A^\dagger\lambda_aA]$.
After some lengthy algebra
the operator   ${\hat g}_{\rm A}^a$ defined in Eq.(\ref{g5d1}) can be
written as:
\bea  {\hat g}_{\rm A}^a &=& -{N_c\over 4} \int  dt\; {d\omega\over 2\pi}
      {d\omega'\over 2\pi}\;  {2\omega E_n + 2 \omega' E_m \over
       (\omega^2+E_n^2)  (\omega^{\prime 2}+E_m^2) }\;
      \exp{(i(\omega-\omega')(t-t_0))} \nn & &
      \int_0^\infty { du \ u} \phi(u)
     \exp{\left(-u[\alpha (\omega^2+E_n^2) + (1-\alpha)
       (\omega^2+E_m^2)]\right)}
        \nn & &
     <n\mid\lambda^c\mid m><m\mid\sigma_3\lambda^b\mid n>
    {\cal T}(\; \Omega_{\rm E}^c(t) D_{ab}(t_0) \;).
      \label{aa20}
\eea
Performing now the $dt$ integration with special care to the
{\it time-ordered} product
${\cal T}(\; \Omega_{\rm E}^c(t) D_{ab}(t_0)\; )$
one gets the relation:
\bea
& &
\int dt \exp{(i(\omega-\omega')(t-t_0))} \;
      {\cal T}[\Omega_{\rm E}^c(t) D_{ab}(t_0)] \nn  &  & =
       {1\over i} \left[  {\rm PP}{1\over\omega-\omega'} + i\pi\delta
        (\omega-\omega') \right]  D_{ab}(t_0)  \Omega_{\rm E}^c
-{1\over i} \left[  {\rm PP}{1\over\omega-\omega'} - i\pi\delta
        (\omega-\omega') \right]   \Omega_{\rm E}^c  D_{ab}(t_0)   \nn
       &  & =  {1\over i} {\rm PP}{1\over\omega-\omega'}
          \left[ D_{ab}(t_0),  \Omega_{\rm E}^c \right]  +
         \pi\delta (\omega-\omega')
         \left\{ D_{ab}(t_0), \Omega_{\rm E}^c \right\}.
       \label{aa21}    \eea
Note that after the time-ordering the angular velocities are
again assumed
to be time-independent in order to perform the $\int dt$ integration.
Last term in \qeq{aa21} vanishes because the
$\delta$-function
makes the integral in \qeq{aa20} odd in $\omega$.  Therefore if the indices
of $\Omega_{\rm E}^c$ and $D_{ab}$ are such that
	$[ D_{ab},  \Omega_{\rm E}^c ] = 0$ \qeq{aa21} gives identically zero.
Evaluating the $\omega,\omega^\prime$ integration finally gives:
\bea  {\hat g}_{\rm A}^a &=& -{N_c\over 4} {i f_{cdb} D_{ad} \over I_{cc} }
      \sum_{m,n}
     <n\mid\lambda^c\mid m><m\mid\sigma_3\lambda^b\mid n>
     {\cal R}_Q (E_n,E_m),
      \label{aa22}
\eea
where  the regularization function  is given by
\beq  {\cal R}_Q (E_n,E_m) =  {1 \over 2 \pi}
      \int {du} \; \phi(u)
       \int_0^1 d\alpha \; { \alpha E_n - (1-\alpha) E_m \over
        \sqrt{\alpha (1-\alpha)} } \; \exp{(-u[\alpha E_n^2
     +(1-\alpha)E_m^2])},      \label{aa23} \eeq
which
 in contrast to the regularization function for the usual moment
of inertia ${\cal R_I}(E_n,E_m)$  or ${\cal R_\beta}(E_n,E_m)$
is antisymmetric with respect to
$E_m$ and $E_n$. The $du$
integration can be performed analytically in the case of
step like regularization functions $\phi(u)=c_i\theta(u-1/\Lambda_i^2)$
and gives
\beq  {\cal R}_Q (E_n,E_m) = c_i
      \int_0^1{ d\alpha  \over 2 \pi}\; { \alpha E_n - (1-\alpha) E_m \over
       \sqrt{\alpha (1-\alpha)}  }     \;
     { \exp{(-[\alpha E_n^2+(1-\alpha)E_m^2]/\Lambda_i^2)}
      \over
      \alpha E_n^2+(1-\alpha)E_m^2   }   . \label{aa24c} \eeq
Using the formula
\beq  \int_0^1 {d \alpha  \over \sqrt{\alpha (1-\alpha) } }
          {1 \over q-\alpha p} = { \pi \over  \sqrt{q(p-q) } }
     ,\ \ {0<p<q}   \label{aa25}  \eeq
the infinite cutoff limit of \qeq{aa24} is given by (p. 219 of
\qref{graryz}):
\beq {\cal R}_Q (E_n,E_m) ={1 \over \mid E_n-E_m \mid }
    { \sign E_n - \sign E_m \over 2 }  \label{aa26}  \eeq
and was used in \qref{ab9} to calculate the $1/N_c$ corrections.
Defining
\beq  Q_{bc} = {N_c\over 4}
    \sum_{m,n}
   <n\mid\lambda^c\mid m><m\mid\sigma_3\lambda^b\mid n>
    {\cal R}_Q (E_n,E_m)
    \label{aa27}
\eeq
the operator ${\hat g}_{\rm A}^a$ can be rewritten as:
\beq   {\hat g}_{\rm A}^a  = -{2i Q_{12} \over I_1} D_{a3}
      -{2i Q_{45} \over I_2} D_{a3}                 .
        \label{aa28}
\eeq

Equation (\ref{aa28}) follows directly from the real part of the Euclidean
effective action given by \qeq{aa1}.
Therefore the corrections described above  have  no counterpart in
the Wess-Zumino term
which follows from the imaginary part of the Euclidean action.
As such they vanish identically in any local mesonic theory like the Skyrme
model for instance.

\section{Comparison with the Gradient Expansion }

In order to check the results of the numerical diagonalization one
should always consult the long wave length expansion of the coefficients
appearing in the expressions for the observables. This technique  is
described at length in \qref{ait}.
It  also clarifies
the question, whether the exact numerical value can be approximated by
the gradient expanded quantities,
or in other words, whether the local mesonic theory like the Skyrme model
for instance, is a good approximation to the NJL-model.

\subsection{The lowest order result from the real part of the EEA}

For the lowest order ($\Omega^0$)  only the
quantity $M_3$, which already exists in  SU(2), contributes to
$g_{\rm A}^{(3)}$.
Its gradient expansion can be found by expanding
$D_0^\dagger~D_0=-\partial^2+M^2+iM\gamma_i\partial_i U(x)$
in \qeq{aa5}
in terms of the gradients  $\partial_i U(x)$.
The result is:
\beq   M_3^{\rm grd} =  {2 \over 3} \int d^3x \left(
                  \sigma \partial_i \pi_i -            \pi_i
               \partial_i \sigma \right). \label{M3grd}  \eeq
\qeq{M3grd} can be rewritten
in terms of the chiral angle $\theta$ and
for $\pi$ and $\sigma$ on the chiral circle
$\sigma(r)=\cos\theta(r)$ and $\pi (r)=\sin\theta(r)$:
\beq  M_3^{\rm grd} = {8 \pi\over 3}f_\pi^2 \int dr\ r^2 \left( \theta^\prime +
       { 2 \sin\theta \cos\theta \over r } \right) . \eeq
For the simplest case of a linear profile $\theta(r)=\pi(1-r/2R)$, it
reduces to:
\beq M_3^{\rm grd} = -{32\pi^2\over 9} f_\pi^2 R^2 \left( 1 - {3\over 2\pi^2}
       \right).  \eeq
This quadratic behaviour of $M_3^{\rm grd}$ is explicitly checked by
using a large $R$ profile function $\theta$  as an input for the quark
wave functions of the exact formula for $M_3$.

Another quantity from the real part of the EEA emerges
due to the presence of a finite $m_s$.
One obtains in leading order:
\beq {\bar R}_{83}^{{\rm grd}}
      =   {1 \over 6 g} \int d^3x \partial_i \pi^i (x),
     \label{beta2}
    \eeq
where we have used  the formula for the normalization
of the kinetic term for mesons, which fixes the cutoff for
the quadraticly divergent integral $I_2(M)$.
As a total divergence the behaviour of this term is
determined just by the asymptotics of the pion field.

Therefore it vanishes for the linear profile discussed above or in
the case of finite $m_{\pi}$, when the profile  decreases  exponentially.
In order to check our numerical calculations we have used
$\theta=2 {\rm arctan}(-R^2/r^2)$.
 Then
$ \overline{R}_{83}^{{\rm grd}} = - (4 \pi /  3 g)  f_\pi R^2 $.
As stated in the text, such  term is not found  in the Skyrme model
and it would actually give a vanishing contribution due to the
asymptotics of the profile.
In the present non-local quark model  ${\bar R}_{83}$ does have a
non-vanishing contribution
(compare with \qtab{tabMR83}).

\subsection{The anomalous terms from the imaginary part of the EEA}

The axial vector current
gets a contribution
 from the imaginary part of the EEA, which  is non-vanishing only in
the SU(3) case. In a local mesonic theory it can be
derived from the Wess-Zumino term \quref{wit83b}.
Here want to show shortly how to derive this contribution from
the non-local EEA of the present NJL model.
Consider a quantity:
\beq  {\rm Im}\; A_i^a (x) = \half\int {\cal D}A(t)  {\delta\over\delta
	 s^a(x)} \Spto \left[  {1\over D}   - {1\over D^\dagger} \right]
		i\gamma_4\gamma_i\gamma_5\lambda^bD_{ab} s^a(x).
\eeq
with $D=\partial_t+H+i\Omega_E$.
Going to the operator form  and using $D^\dagger D$ one can write
\beq   {\rm Im} {\hat A}_i^a (x) = \half{\delta\over\delta s^a(x)}
      \Spto  {1\over D^\dagger D}
    \left[D^\dagger i\gamma_4\gamma_i\gamma_5\lambda^bD_{ab}
    - i\gamma_4\gamma_i\gamma_5\lambda^bD_{ab}  D
      \right] s^a(x)
\eeq
Expanding $D^\dagger~D$  again in terms of the gradients
leads after some laborious algebra:
\beq M^{\rm grd}_{44}   = {\Nc\over 16 \pi^2} {1\over f_\pi^3}
                \epsilon_{0\mu\nu 3} \epsilon_{3ab}
      \int d^3 x
    \partial_\mu \pi^a(x) \partial_\nu \pi^b(x)
      \sigma(x) .    \label{M4grd}
\eeq
In the case of the {\it hedgehog} Ansatz i.e.
$\sigma(r)=\cos\theta(r)$ and $\pi (r)=\sin\theta(r)$
 Eq.(\ref{M4grd}) reduces to:
\beq  M^{\rm grd}_{44}   = - {\Nc\over 6\pi}     {1\over f_\pi^3}
      \int dr r  \partial_r \sigma (r) \pi(r)^2
       =  {\Nc\over  6\pi}
      \int dr r  \theta^\prime (r) \sin^3 \theta (r).
\eeq
For the linear
profile $\theta(r)=\pi(1- {r/2R})$ we obtain
a compact expression:
\beq M^{\rm grd}_{44}   = - {2\over 3 \pi }  R .  \eeq
This linear behaviour for large size chiral fields can be seen
in Fig. 7.

\subsection{The $\Omega^1$-terms from the real  part of the EEA}

In this Appendix we derive the gradient expansion for the {\it non-local}
terms ($Q_{ab}$). The formulae below are given only for this part
of the axial vector current operator ${\hat g}_{\rm A}^a$:
\bea    {\hat g}_{\rm A}^a  & = & - \half \Tr_{\gamma,\tau,c} \int d^3x\ dt
            <{\vec
           x},t_0\mid
          {1\over \partial_t + H}\mid t>
          <t\mid{1\over \partial_t + H}
        \gamma_0\gamma_i\gamma_5\lambda_b \mid{\vec x},t_0> \nn
 & & ~~~~~~~~~~~~~~~~~
{\cal T}\left[\Omega_c(t) D_{ab}(t_0)\right],
\eea
where the regularization is neglected here for simplicity.
Inserting eigenstates of  $\partial_t$ and $H$ and using
\qeq{aa21} we can define
\beq  {\hat g}_{\rm A}^a = \left[ {X_{12}^3 \over I_1}    +
                       {X_{45}^3 \over I_2}  \right]  D_{a3},
\eeq
where the X-quantities can be calculated from:
\beq  X_{bc}^i = \Tr \int {d\omega\over 2\pi}  {d\omega'\over 2\pi}
      {\rm PP}{1\over \omega-\omega^\prime} <{\vec x}\mid
      {1\over-i\omega+H}\lambda_c{1\over-i\omega'+H}
      \gamma_0\gamma_i\gamma_5\lambda_b \mid {\vec x}>.
\eeq
Then the recipy is  to multiply  denominators
and numerators by the hermitian
conjugate of the denominators and recover
$H^2=-\partial_i^2+M^2+iM\gamma_i\partial_iU(x)$
in denominators, which can be expanded
in terms of the gradients.
Then
these expressions can be straightforwardly simplified
to the pure
SU(2) quantity:
\bea   X_{12}^3   &=&
        -{\Nc M\over 48\pi} \int d^3x
         {1\over f_\pi^2} \left( \pi^i\partial_i\sigma +
           \sigma\partial_i \pi^i \right)  \nn
     & &   +{\Nc M\over 16\pi} \int d^3x
         {1\over f_\pi^2} \left(
           \sigma\partial_i \pi^i
 -  \pi^i\partial_i\sigma  \right)  \nn
    &=& -{\Nc M\over 4} \int dr r^2  \left(
         \theta' + {\sin  2\theta\over r}  \right) \nn
 & &        -{\Nc M\over 12}  \int dr r^2  \left(
         \theta'\cos~2\theta + {\sin  2\theta\over r}  \right)
       \label{b18}
\eea
and the   pure SU(3) quantity
\bea   X_{45}^3   &=&
            {\Nc M\over 32\pi} \int d^3x
         {1\over f_\pi^2} \left(
           \sigma\partial_i \pi^i  -  \pi^i\partial_i\sigma  \right)
		\nn
 &=& -{\Nc M\over 8} \int dr r^2  \left(
      \theta' + {\sin  2\theta\over r}  \right),
       \label{b19}
\eea
where the first line for $X_{12}^3$  is a total divergence and
vanishes for
chiral fields, which  vanish  at least as  ${1/ r^2}$ for
$r\rightarrow\infty$.
Assuming physical profiles, which vanish exponentially with the pion
mass, the axial vector current operator can be written  as
\beq
     {\hat g}_{\rm A}^a =  \int d^3r \; r^2 \left(
      \theta' + {\sin  2\theta\over r} \right)
		\left[ {8\pi\over 3} f_\pi^2 + {M\over 4I_1} +
      {M\over 8I_2} \right] \; D_{a3}.
       \label{b20}
\eeq
Note that $I_1,I_2\sim~\Nc$, such that the last two terms in
\qeq{b20} represent a $1/\Nc$ correction.
Therefore  \qeq{b20} resembles very much the  result of
Dashen and Manohar  \cite{bolo:dama2,bolo:dama1}, which states that
the $1/\Nc$ corrections to the axial current lead only to a
renormalization of $g_{\rm A}$. Or in other words, the ratio of different
coupling constants  has {\it no} $1/\Nc$ correction.

\vfill\eject
\begin{figure}
\caption{Constituent quark mass $M$ as a function of  of
$\ms$ such that the octet-decuplet splitting is reproduced.
}
\end{figure}
\begin{figure}
\caption{The model deviations (i.e. {\it theory-experiment})
for octet and decuplet baryons
as
functions of
$\ms$ for the optimal constituent quark mass $M$ of Fig.1}
\end{figure}
\begin{figure}
\caption{The theoretical deviations from the experimental values
of the $\Sigma$ and $\Lambda$ particle for $M=419$ MeV.
Compared are a perturbative treatment (pert1) of the wave functions in
zeroth oder, a linear corrections of order $m_s$ (pert2) and an exact
diagonalization  (YA), proposed by Yabu and Ando [37].
}
\end{figure}
\begin{figure}
\caption{The coefficients $c_{\bar 10}$ and $c_{27}$ of the higher
representations
$\bf{\bar 10}$
and $\bf 27$ of the proton as a function of the constituent quark mass
M. The strange current quark mass  is chosen as $m_s=180MeV$
}
\end{figure}
\begin{figure}
\caption{The anomalous  quantity $M_{44}$ compared with the leading term
of a gradient expansion, which comes exactly from the Wess-Zumino action
for $SU(3)$ pseudoscalar fields.
This is done for  a fixed linear  profile in
dependence of the radius R for $M=372MeV$.
}
\end{figure}
\begin{figure}
\caption{The $g_A^0$, $g_A^3$ and $g_A^{(8)}$  are shown for
selfconsistent  chiral fields in dependence of the constituent quark
mass. The strange current quark mass is chosen as $m_s=180MeV$,
according to a best fit  to the hyperon spectra.
}
\end{figure}
\begin{figure}
\caption{The regularization function  ${\cal R}_Q(E_n,E_m)$
for the time-ordered expressions  (reg)
for fixed $E_n$ and $M=400MeV$   in dependence of $E_m$.
This is compared to the function (noreg), which is obtained
in the infinite cutoff limit.
}
\end{figure}
\vfill\eject

\pagebreak
\end{document}